\documentclass{aa}
\usepackage{longtable}
\usepackage{graphicx}

\usepackage{natbib,twoopt}
\usepackage[breaklinks=true]{hyperref} %% to avoid \citeads line fills
\bibpunct{(}{)}{;}{a}{}{,}             %% natbib format for A&A and ApJ
\makeatletter
  \newcommandtwoopt{\citeads}[3][][]{\href{http://adsabs.harvard.edu/abs/#3}%
    {\def\hyper@linkstart##1##2{}%
     \let\hyper@linkend\@empty\citealp[#1][#2]{#3}}}
  \newcommandtwoopt{\citepads}[3][][]{\href{http://adsabs.harvard.edu/abs/#3}%
    {\def\hyper@linkstart##1##2{}%
     \let\hyper@linkend\@empty\citep[#1][#2]{#3}}}
  \newcommandtwoopt{\citetads}[3][][]{\href{http://adsabs.harvard.edu/abs/#3}%
    {\def\hyper@linkstart##1##2{}%
     \let\hyper@linkend\@empty\citet[#1][#2]{#3}}}
  \newcommandtwoopt{\citeyearads}[3][][]%
    {\href{http://adsabs.harvard.edu/abs/#3}
    {\def\hyper@linkstart##1##2{}%
     \let\hyper@linkend\@empty\citeyear[#1][#2]{#3}}}
\makeatother

\newcommand{\um}{\hbox{$\mu$m}}
\newcommand{\giv}{\;|\;}

\newcommand{\Lx}{$L_\mathrm{X}$}

\newcommand{\mstel}{\mathcal{M}_*}
\newcommand{\Mstel}{$\mstel$}

\newcommand{\ledd}{\lambda_{\mathrm{Edd}}} 
\newcommand{\pledd}{p(\lambda_{\mathrm{Edd}})} 
\newcommand{\dd}{\mathrm{d}}
\authorrunning{Wang et al.}
\titlerunning{The dependence of AGN accretion on host colors at $0.5 < z < 2.5$ }

\begin{document} 

\title{\textbf{AGN-Host Connection at $0.5 < z < 2.5$: A rapid evolution of AGN fraction in red galaxies during the last 10 Gyr}}
\author{Tao Wang\inst{1,2}
      \and D. Elbaz\inst{1}
      \and D. M.  Alexander \inst{3}
      \and Y. Q. Xue \inst{4}
      \and J. M. Gabor\inst{1}
      \and S. Juneau\inst{1}
      \and C. Schreiber\inst{1,5}
      \and X-Z. Zheng \inst{6}
      \and S. Wuyts\inst{7}
      \and Y. Shi\inst{2}
      \and E. Daddi \inst{2}
      \and X-W. Shu\inst{8}
      \and G-W. FANG\inst{9}
      \and J-S. Huang \inst{10,11}
      \and B. Luo\inst{2}
      \and Q-S. Gu\inst{2}   
      }
\institute{Laboratoire AIM-Paris-Saclay, CEA/DSM/Irfu, CNRS, Universit\'e Paris Diderot, Saclay, pt courrier 131, 91191 Gif-sur-Yvette, France 
\\
\email{taowang@nju.edu.cn}
\and Key Laboratory of Modern Astronomy and Astrophysics in Ministry of Education, School of Astronomy \& Space Science, Nanjing University, Nanjing 210093, China
\and Department of Physics, Durham University, South Road, Durham DH1 3LE, UK
\and CAS Key Laboratory for Researches in Galaxies and Cosmology, Department of Astronomy, University of Science and Technology of China, Chinese Academy of Sciences, Hefei, Anhui 230026, China
\and Leiden Observatory, Leiden University, NL-2300 RA Leiden, The Netherlands
\and Purple Mountain Observatory, Chinese Academy of Sciences, 2 West-Beijing Road, Nanjing 210008, China
\and Department of Physics, University of Bath, Claverton Down, Bath, BA2 7AY, UK
\and Department of Physics, Anhui Normal University, Wuhu, Anhui,241000, China
\and Institute for Astronomy and History of Science and Technology, Dali University, Dali 671003, China
\and National Astronomical Observatories of China, Chinese Academy of Sciences, Beijing 100012, China
\and Harvard-Smithsonian Center for Astrophysics, 60 Garden Str., Cambridge, MA02138, USA
}

\abstract{
We explore the dependence of the incidence of moderate-luminosity ( $L_{0.5-8~\mathrm{keV}} = 10^{41.9-43.7}$ erg s$^{-1}$) Active Galactic Nuclei (AGNs) and the distribution of their accretion rates on host color at $0.5 < z < 2.5$. Based on the deepest X-ray and UV-to-far-infrared data in the two The Great Observatories Origins Deep SurveyGOODS fields, we identify 221 AGNs within a mass-complete parent galaxy sample down to $M_{*} > 10^{10} M_{\odot}$. 
We use extinction-corrected rest-frame $U - V$ colors to divide both AGN hosts and non-AGN galaxies into red sequence (red), green valley (green), and blue cloud (blue) populations. 
We find that the fraction of galaxies hosting an AGN at fixed X-ray luminosity increases with  stellar mass and redshift for all the three galaxy populations, independent of their colors. However, both the AGN fraction at fixed stellar mass and its evolution with redshift are clearly dependent on host colors. Most notably, red galaxies have the lowest AGN fraction ($\sim$5\%) at $z \sim 1$ yet with most rapid evolution with redshift, increasing by a factor of $\sim$ 5 (24\%) at $z \sim 2$. Green galaxies exhibit the highest AGN fraction across all redshifts, which is most pronounced at $z\sim 2$ with more than half of them hosting an AGN at $M_{*} > 10^{10.6} M_{\odot}$. Together with the high AGN fraction in red galaxies at $z \sim 2$, this indicates that (X-ray) AGNs could be important in both transforming (quenching) star-forming galaxies into quiescent ones and subsequently maintaining their quiescence at high redshift. Furthermore, consistent with previous studies at lower redshifts, we show that the probability of hosting an AGN for the total galaxy population can be characterized by a universal Eddington ratio (as approximated by $L_{X}/M_{*}$) distribution ($p(\lambda_{\mathrm{Edd}}) \sim \lambda_{\mathrm{Edd}}^{-0.4}$), which is independent on host mass. 
Yet consistent with their different AGN fractions, galaxies with different colors appear to also have different $p(\lambda_{\mathrm{Edd}})$ with red galaxies exhibiting more rapid redshift evolution compared with that for green and blue galaxies. Evidence for a steeper power-law distribution of $p(\lambda_{\mathrm{Edd}})$ in red galaxies ($p(\lambda_{\mathrm{Edd}}) \sim \lambda_{\mathrm{Edd}}^{-0.6}$) is also presented, though larger samples are needed to confirm. These results suggest that the AGN accretion or the growth of supermassive black holes is related to their host properties, and may also influence their hosts in a different mode dependent on the host color.}

\maketitle

\section{Introduction} 
The evolution of galaxies and central black holes is closely related. Observations have revealed a tight correlation between the mass of supermassive black holes (SMBH) and the velocity dispersion of the host galaxy bulge in the local universe, that is, $M_{BH} - \sigma$ relation,
indicating that the growth of the SMBH may be intimately tied to the build-up of the host galaxy \citep{Magorrian:1998,Ferrarese:2000,Gebhardt:2000,Alexander:2012,Kormendy:2013}. It is also found that the evolution of the volume density of SMBH accretion rate is very similar to that of the cosmic star formation rate up to $z \sim 3$ \citep{Heckman:2004,Silverman:2008b,Aird:2010,Dunlop:2011,Mullaney:2012b}. 
On the other hand, theoretical models propose that feedback from rapidly accreting SMBHs in the active galactic nucleus (AGN) phase is required to quench star formation in massive star-forming galaxies (``Quasar-mode'') and keep quiescent galaxies red and dead (``Radio-mode'')\citep{Silk:1998,di_matteo:2005,Best:2005,Croton:2006}. However, observationally it remains unclear how the fueling of SMBHs is related to star formation in the host and in which galaxies the AGN feedback actually occurs.  
Central to our understanding of these questions is to determine simultaneously both accretion states of  AGNs and physical properties of host galaxies across cosmic time. 
In particular, studying AGNs and their hosts at $z \sim 0.5-2.5$ is of great importance, as it is then when the majority of the growth of SMBHs and of the stellar components in galaxies occurrs \citep{Silverman:2008b}. 

During the last decades, numerous studies of AGNs and their hosts in the local Universe have provided us a clear picture of the fossil record of galaxy/SMBH coevolution, as well as informative insights into the processes that govern the coevolution \citep[see a recent review by][]{Heckman:2014}.
Based on a large sample of optically-selected narrow-line AGNs in SDSS, most AGNs are found to be hosted in massive galaxies ($M \gtrsim 10^{10} M_{\odot}$) with young stellar populations, and tend to reside in the green valley and the top end of the blue cloud in the color-magnitude diagram \citep{Kauffmann:2003_agn,Martin:2007,Wild:2007,Hernan-Caballero:2014}. These results lead to arguments that AGNs may play an important role in transforming blue star-forming galaxies to red and dead galaxies.

More detailed characterization of  AGN host galaxies further reveals the dependence of AGN activity on various host properties. For instance, by sorting galaxies into different morphological types, \cite{Schawinski:2009b} show that the probability of hosting an AGN as well as the role of AGN in shaping the evolution of the host galaxy strongly depend on host morphologies. Similarly, by dividing AGN hosts into passive and star-forming galaxies, \cite{Kauffmann:2009} reveal that the two types of hosts have distinct Eddington ratio distributions, suggesting that the accretion of AGNs strongly depends on the star-formation status of their hosts. Moreover, with a hard X-ray-selected AGN sample, that is, the $Swift$ BAT AGN sample, \cite{Koss:2011} show that these AGNs tend to show bluer colors and a higher fraction of spirals and mergers compared to inactive galaxies and optically-selected AGNs from SDSS. Although it still remains unclear what causes these differences, their findings also suggest that the host galaxy morphologies/colors are somehow related to the activation and fueling of AGNs.

At higher redshifts, however, the relation between the activation and fueling of AGNs and their host galaxies is more elusive. 
Benefiting from recent deep X-ray surveys, which allow efficient identifications of typical AGNs to progressively higher redshift, a number of studies have provided instructive insight into the AGN-host relation but have often yielded discrepant results.
Several works show that galaxies hosting an X-ray AGN at $z \sim 1$ tend to have intermediate colors, placing them in the green valley \citep{Nandra:2007,Bundy:2008,Georgakakis:2008,Silverman:2008b,Hickox:2009,Schawinski:2009c,Treister:2009}, but some later works show that this may be primarily due to mass-selection effects, and argue that when comparing to mass-selected samples, either there is no preference of AGN hosts in color~\citep{Xue:2010} or a preference of AGN hosts for blue/star-forming galaxies~\citep{Aird:2012,Rosario:2013b}.  
On the other hand, based on an X-ray stacking analysis, \cite{Olsen:2013} found that at $z \sim 2$, quiescent galaxies appear to have a higher (most likely low-luminosity) AGN fraction than star-forming galaxies. How to reconcile these different results at high redshifts and obtain a coherent picture connecting results at high and low redshifts remains one of the main challenges in understanding the AGN-host connection.

There are several effects that may potentially cause the aforementioned discrepancies in the AGN-host connection, specifically, how the activation and fueling of AGNs depend on host-galaxy properties, such as, mass, color, and star formation rate. 
One is the mass-selection effect: AGNs are more easily detected in massive galaxies (with more massive black holes) which tend to have redder colors \citep{Silverman:2009,Xue:2010,Aird:2012}. Thus, a carefully selected mass-matched control sample is required when comparing the relative prevalence of AGN hosts among parent galaxies. Another effect is the impact of dust-reddening on host colors: both quiescent galaxies with intrinsic evolved stellar populations and star-forming galaxies reddened by dust can appear on the red sequence/green valley; and the observed colors become a poorer tracer of the actual level of star formation at progressively higher redshifts, when dusty star-forming galaxies are more prevalent \citep{Cardamone:2010, Brammer:2009}. Without taking this into account, it would be difficult to characterize the true dependence of AGNs on host star formation, as well as to make a fair comparison between studies at different redshift. 

\begin{figure}[t!]
\begin{center}
\includegraphics*[trim=-10 -30 0 20,width=8.8cm]{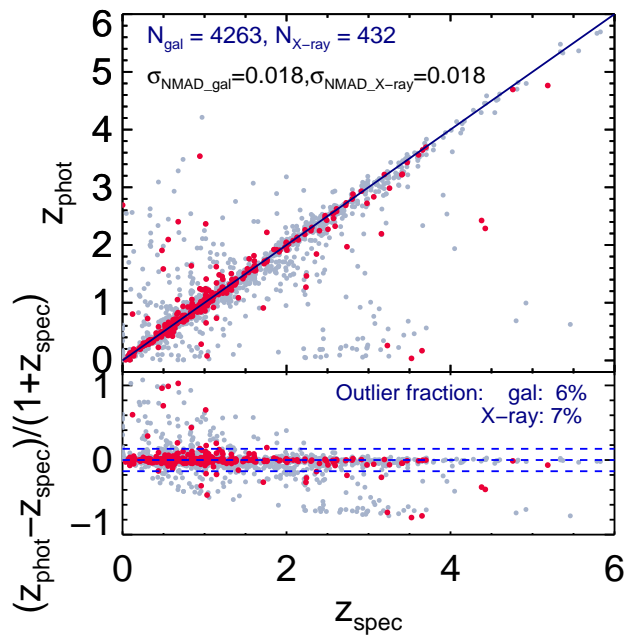}\caption{Spectroscopic redshift ($z_{spec}$) versus photometric redshift ($z_{phot}$) for all galaxies in GOODS-South and GOODS-North with available $z_{spec}$. Galaxies that are detected in X-ray are denoted as red filled circles. }
\label{Fig:zs_zp}
\end{center}
\end{figure}

In this work, we explore the dependence of AGN activity on properties of their hosts at  $0.5 < z < 2.5$. We mainly focus on two  questions: whether AGNs are preferentially found in certain types of host galaxies, and whether the growth of SMBHs is related to the properties of their hosts.
The main difference between this study and most previous studies is that we dissect both AGN hosts and non-AGN galaxies by de-reddened rest-frame colors~\citep[][also see, e.g., ]{Cardamone:2010,Cimatti:2013}. We select our galaxy and AGN samples in the GOODS-North and GOODS-South fields, which have the deepest X-ray observations to date as well as rich ancillary multi-wavelength data. In particular, the new near-infrared data from the Cosmic Assembly Near-infrared Deep Extragalactic Legacy Survey (CANDELS,\citealt{Grogin:2011,Koekemoer:2011}) and the 3D-HST survey \citep{Brammer:2012,Skelton:2014} 
permit more accurate estimates of photometric redshifts, stellar masses, and rest-frame colors for galaxies in the sample. 
This allows us to study typical AGNs with $10^{41.9} < L_{X} < 10^{43.7}$ erg $s^{-1}$(Seyferts like)  in a mass-selected galaxy sample down to $M_{*} = 10^{10.0} M_{\odot}$ at $z =2.5$. 

The outline of this paper is as follows. The selection and properties of the AGN and parent galaxy samples are described in Section 2. We describe how we derive de-reddened galaxy colors, and study the dependence of AGN fraction on galaxy color in Section 3. We then explore how the properties of AGN vary as a function of galaxy color in Section 4. In Section 5 we study the dependence of the Eddington ratio distribution on host colors and explore the physical origins of this dependence in Section 6. We then compare our results to previous studies and discuss the effects of sample selection in Section 7. We summarize our main results in Section 8. Throughout the paper, we assume an $\Omega_\Lambda = 0.7$, $\Omega_{M} = 0.3$, and $H_{0}$ = 70 km s$^{-1}$ Mpc$^{-1}$ cosmology. All magnitudes are in the AB system.

\section{Data}
\subsection{Source catalogs \label{data}}

For GOODS-South, we utilize the official CANDELS HST/WFC3 $H$-band selected multi-wavelength catalog \citep{GuoY:2013} as our base catalog. The new HST/WFC3 observations from the ERS  \citep{Windhorst:2011} and CANDELS GOODS surveys reach 5 $\sigma$ limiting magnitude $H \gtrsim 27.4$ \citep{Grogin:2011,Koekemoer:2011}. Since the resolution of the images significantly changes from the optical to the IR band, an object template-fitting software dubbed TFIT \citep{Laidler:2007} is used to robustly measure the photometry of objects in all U-to-8.0 $\um$ bands, including $U$ (VLT/VIMOS), $BViz$ (HST/ACS), $F098M, F105W, F125W, F160W$ (HST/WFC3), and 3.6, 4.5, 5.8, 8.0 $\um$ (Spitzer/IRAC) band.

For GOODS-North, we use a HST/WFC3 NIR-selected multi-wavelength catalog  presented in \cite{Skelton:2014}, which is based on combined  F125W+F140W+F160W images. The ultra-deep near-infrared imaging data reach  5 $\sigma$ total limiting magnitudes of  $\sim$ 27 mag in the F160W band. The multi-wavelength catalog also includes point spread function (PSF)-matched photometry in $U$ (KPNO/MOSAIC), $BViz$ (HST/ACS), $JHK_{s}$ (Subaru/MOIRCS \citep{Yamada:2009,Kajisawa:2010}), and 3.6, 4.5, 5.8, 8.0 $\um$ ({\it Spitzer}/IRAC) bands. Both GOODS-South and GOODS-North catalogs are primarily based on the CANDELS survey, which reach similar depth and are complete for galaxies with $M_{*} > 10^{9} M_{\odot}$ at $z \sim 2$~\citep{Grogin:2011}. Comparisons between the 3H-HST and CANDELS catalog in GOODS-North (Barro et al., in preparation) show that the photometry agrees reasonably well, particularly at the HST bands.

\begin{figure}[t!]
\begin{center}
\includegraphics[trim = -20 0 0 0,width=8.8cm]{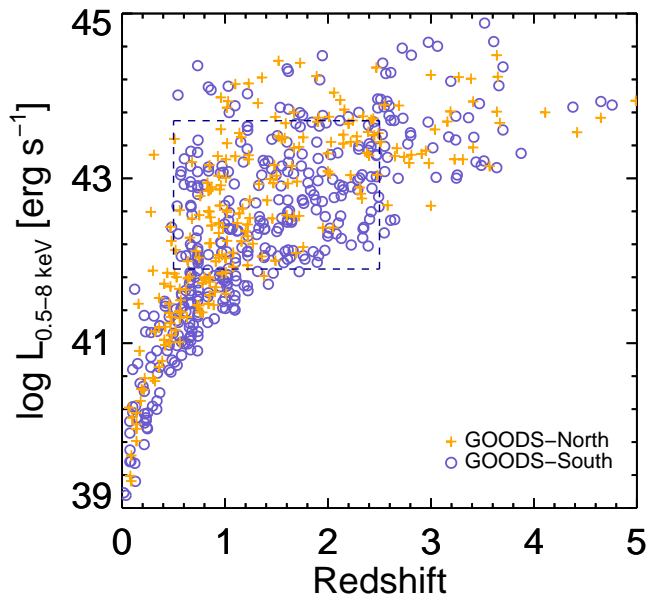}
\caption{X-ray luminosity vs. redshift for the X-ray detected sub-sample of galaxies in CDF-N (yellow crosses) and CDF-S (blue circles). We restrict our sample to moderate-luminosity AGNs with 41.9 $<$ log $L_{X}$ $<$ 43.7 erg $s^{-1}$, and at 0.5 $< z <$ 2.5. 
\label{Fig:lx_z}}
\end{center}
\end{figure}

The GOODS-North and GOODS-South fields have the deepest available X-ray data. We made use of the 2Ms X-ray source catalog in \cite{Alexander:2003} and the 4Ms main catalog in \cite{Xue:2011} for GOODS-North and GODOS-South, respectively. The on-axis 0.5$-8$ kev sensitivity limits reach $\sim 7.1 \times 10^{-17}$ erg cm$^{-2}$ s$^{-1}$ and $3.2 \times 10^{-17}$ erg cm$^{-2}$ s$^{-1}$ for GOODS-North and GOODS-South, respectively.  

We cross match the optical/NIR catalogs and the X-ray source catalogs using a searching radius of 1.5\arcsec. In total, we identify 400 (out of 403) X-ray sources with H-band counterparts in GOODS-South and 297 (out of 313) sources in GOODS-North respectively. 
We also include $Spitzer/MIPS$ 24$\mu$m and $Herschel$/PACS data by cross-matching the NIR-selected catalog with 24 $\mu$m-selected catalogs in both GOODS-North and GOODS-South \citep{Magnelli:2013}. These 24 $\mu$m catalogs also include photometry at PACS 100 $\mu$m and 160 $\mu$m from the PACS Evolutionary Probe (PEP;\citealt{Lutz:2011}) and GOODS-Herschel key programs \citep{Elbaz:2011}, which reach typical 5$\sigma$ depths of 21~$\mu$Jy, 1.1 mJy and 2.7 mJy at 24 $\mu$m, 100 $\mu$m and 160 $\mu$m, respectively.

\subsection{Sample construction}
 \label{Sec:properties}
For all galaxies in the source catalogs, we adopt high-quality spectroscopic redshifts if available \citep[and references therein]{Luo:2008, Xue:2010,Dahlen:2013}; otherwise, we use photometric redshifts (photo-z's) from \cite{Hsu:2014} for GOODS-South and  \cite{Skelton:2014} for GOODS-North. In particular, the photo-z's estimates for X-ray sources in GOODS-South from \cite{Hsu:2014} are based on a library of AGN/galaxy hybrid templates. For GOODS-North, the photo-z's estimates in \cite{Skelton:2014} are derived with $EAZY$ \citep{Brammer:2008}, which only includes galaxy templates. Recently \cite{YangG:2014} also derive photo-z's for X-ray sources in Hawaii-Hubble Deep Field-North (H-HDF-N, which includes the GOODS-North region) by adding additional AGN templates in their template libraries. We do not find significant improvements compared to the photo-z's from \cite{Skelton:2014}, which may be partly due to the fact that \cite{YangG:2014} used a much shallower data set. We thus chose to use the photo-z's from \cite{Skelton:2014}, which used data sets of similar depth to GOODS-South. 

We show the quality of the photo-z's fitting combining the two fields in Figure~\ref{Fig:zs_zp}. The normalized median absolute
deviation ($\sigma_\textsc{nmad} = 1.48 \times \mathrm{median}\left( \left| \frac{\Delta z-\mathrm{median}(\Delta z)}{1+z_{spec}} \right| \right)$,~\citealt{Brammer:2008}) of $\Delta z = z_{phot}-z_{spec}$ reaches $\sigma _\textsc{NMAD}$ $\sim$ 0.018 for both X-ray detected and non-detected galaxies. The fraction of catastrophic outliers (i.e., objects with $|\Delta z|/(1+ z_{spec}) > 0.15$) is slightly higher for X-ray sources, which is 7\% compared to 6\% for non-X-ray sources. Note that $\sim$ 62\% (432/697) of X-ray sources in our sample have reliable spectroscopic redshifts. 

We then employ FAST \citep{Kriek:2009a} to estimate the
stellar mass, star-formation rate, and dust content ($A_{V}$) for each galaxy. We construct galaxy templates from the \cite{Maraston:2005} stellar population synthesis models with a \cite{Kroupa:2001} initial mass function and solar metallicity, assuming exponentially declining star-formation histories (SFHs) with the $e$-folding time $\tau \sim 0.1-10$ Gyr. We avoid shorter e-folding time scales, for example, $\tau < 0.1$ Gyr, because this usually leads to systematic offsets in SFR compared to measurements of SFR$_{UV+IR}$ as shown in \cite{Wuyts:2011a}. We allow the galaxies to be attenuated within $A_{V}$ = 0 $-$ 4 with the reddening following the \cite{Calzetti:2000} law. To avoid likely contamination of PAH and AGN emission, we exclude the two longer IRAC bands during the SED fitting. 
We have also derived the stellar mass using \cite{Bruzual:2003} models and \cite{Chabrier:2003} IMF, and found that the stellar mass would be, on average, $\sim$0.1 dex higher.

We calculate the rest-frame 0.5$-$8 keV X-ray luminosity ($L_{X}$) for X-ray sources in the two fields assuming a power-law with $\Gamma = 1.8$. The distribution of X-ray luminosity of $Chandra$ sources in the two fields as a function of redshift is shown in Fig~\ref{Fig:lx_z}. We select a moderate-luminosity AGN sample with
$10^{41.9} < L_{X} < 10^{43.7}$ erg s$^{-1}$ at $0.5 < z < 2.5$. The lower luminosity limit is to ensure that X-ray emission is mainly due to an AGN instead of star formation, while the upper limit is to ensure that the optical and near-infrared emission is primarily due to host galaxies \citep{Silverman:2009,Xue:2010}. We do not exclude Type I AGNs in the sample since the spectroscopic coverage is far from homogenous between the two fields and also the fraction of broad-line AGNs within the luminosity range studied here is relatively small. For instance, \cite{Barger:2015} shows that the fraction of Type I AGNs in the range $L_{X} < 10^{43.7}$ erg s$^{-1}$ up to $z \sim 4$ is $\lesssim 10\%$ (and increases substantially above this luminosity).

\begin{figure*}[th!]
\includegraphics[trim = -20 -10 60 10, angle=0,width=0.49\textwidth]{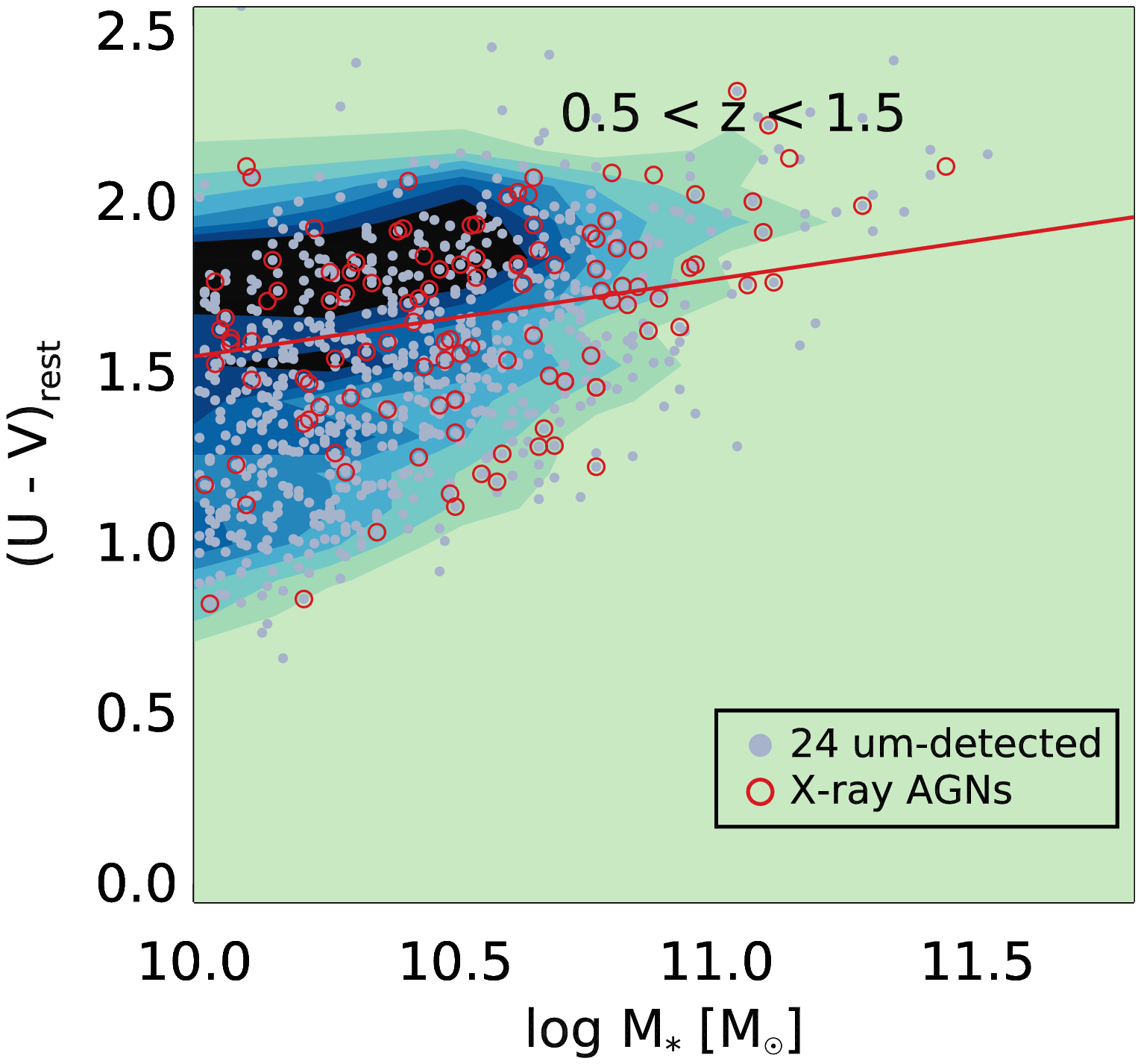}
\includegraphics[trim = -20 -10 60 10, angle=0,width=0.49\textwidth]{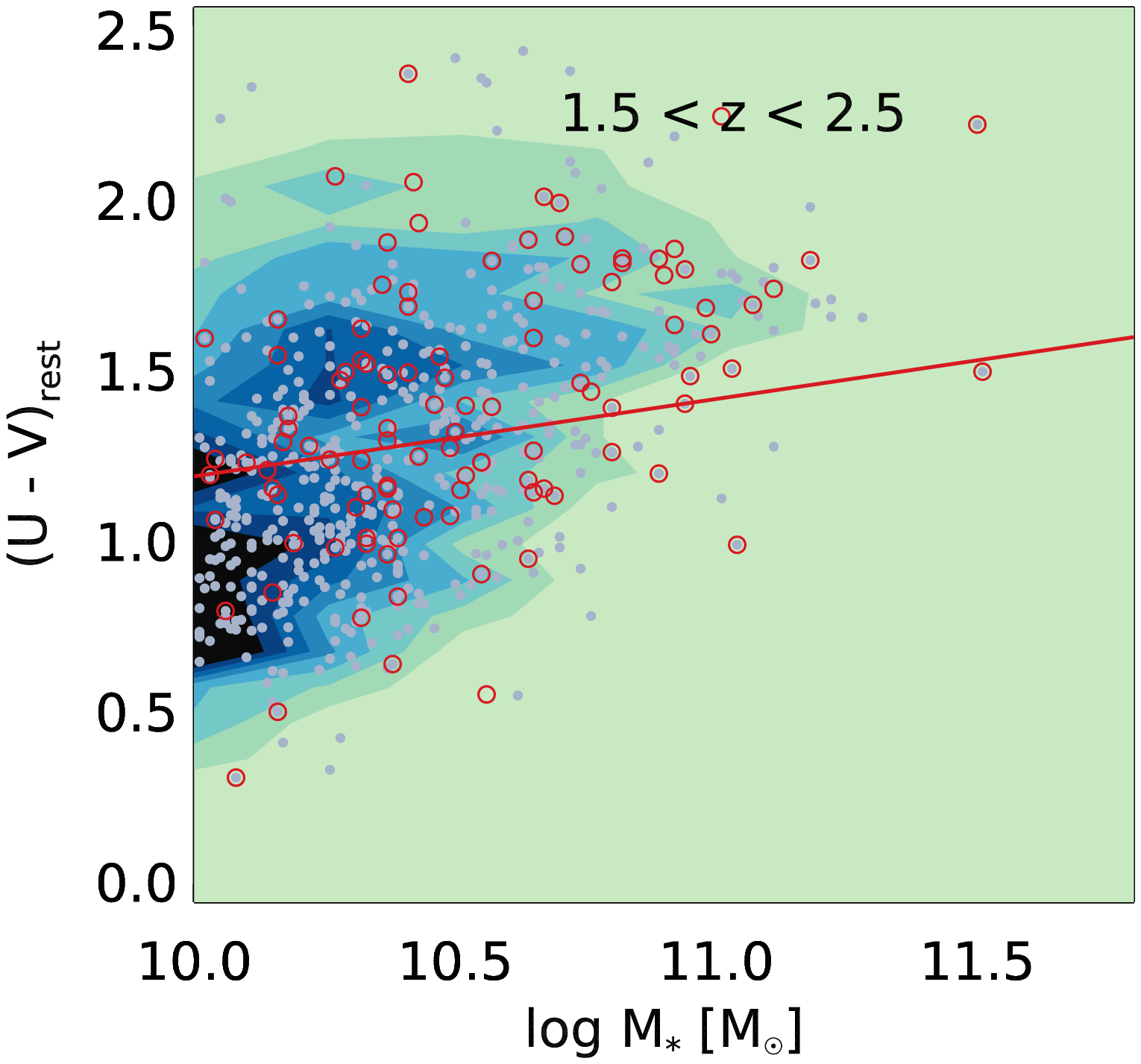}\\
\vspace{0.2in}
\includegraphics[trim = -20 -10 60 10, angle=0,width=0.49\textwidth]{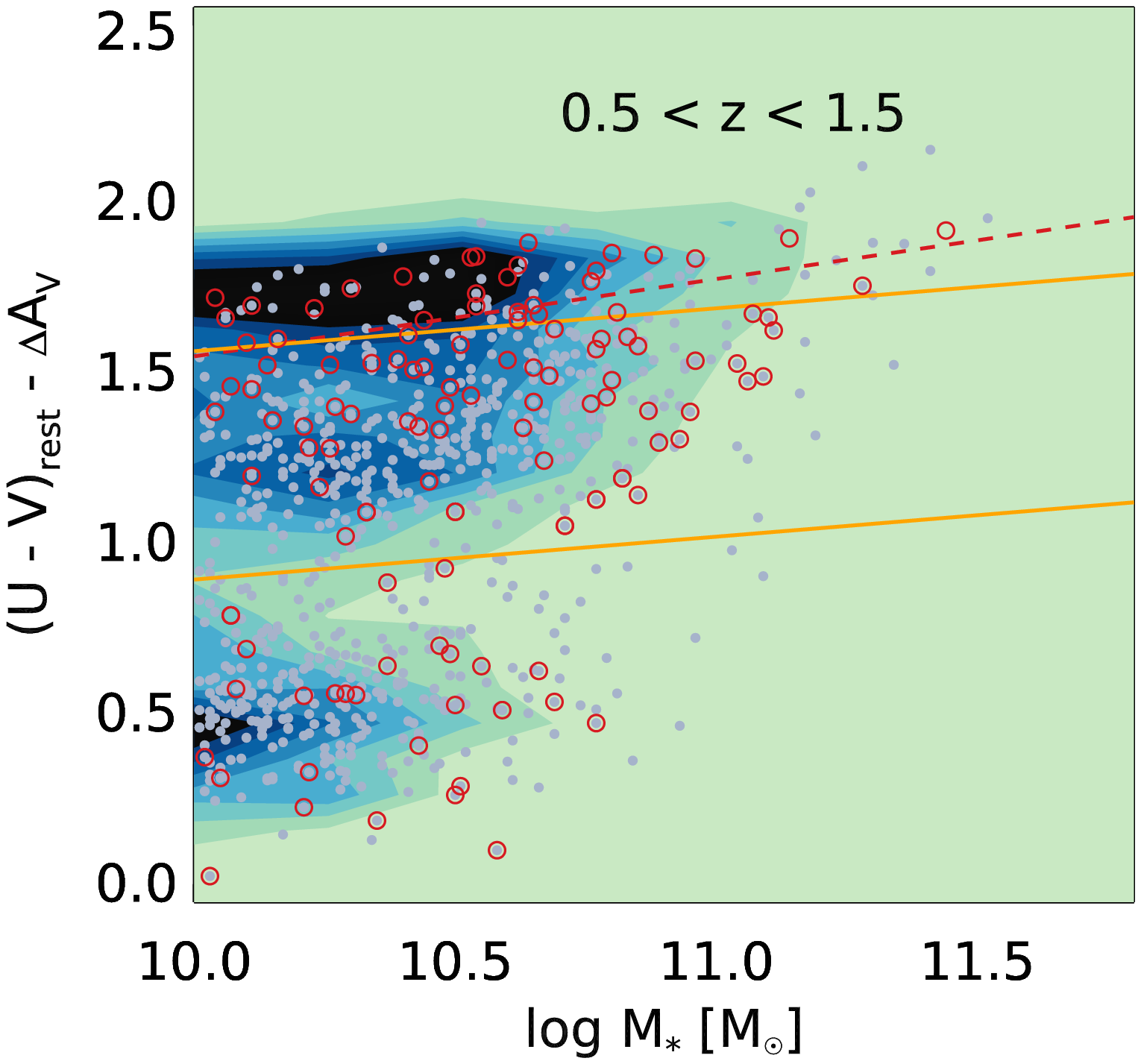}
\includegraphics[trim = -20 -10 60 10, angle=0,width=0.49\textwidth]{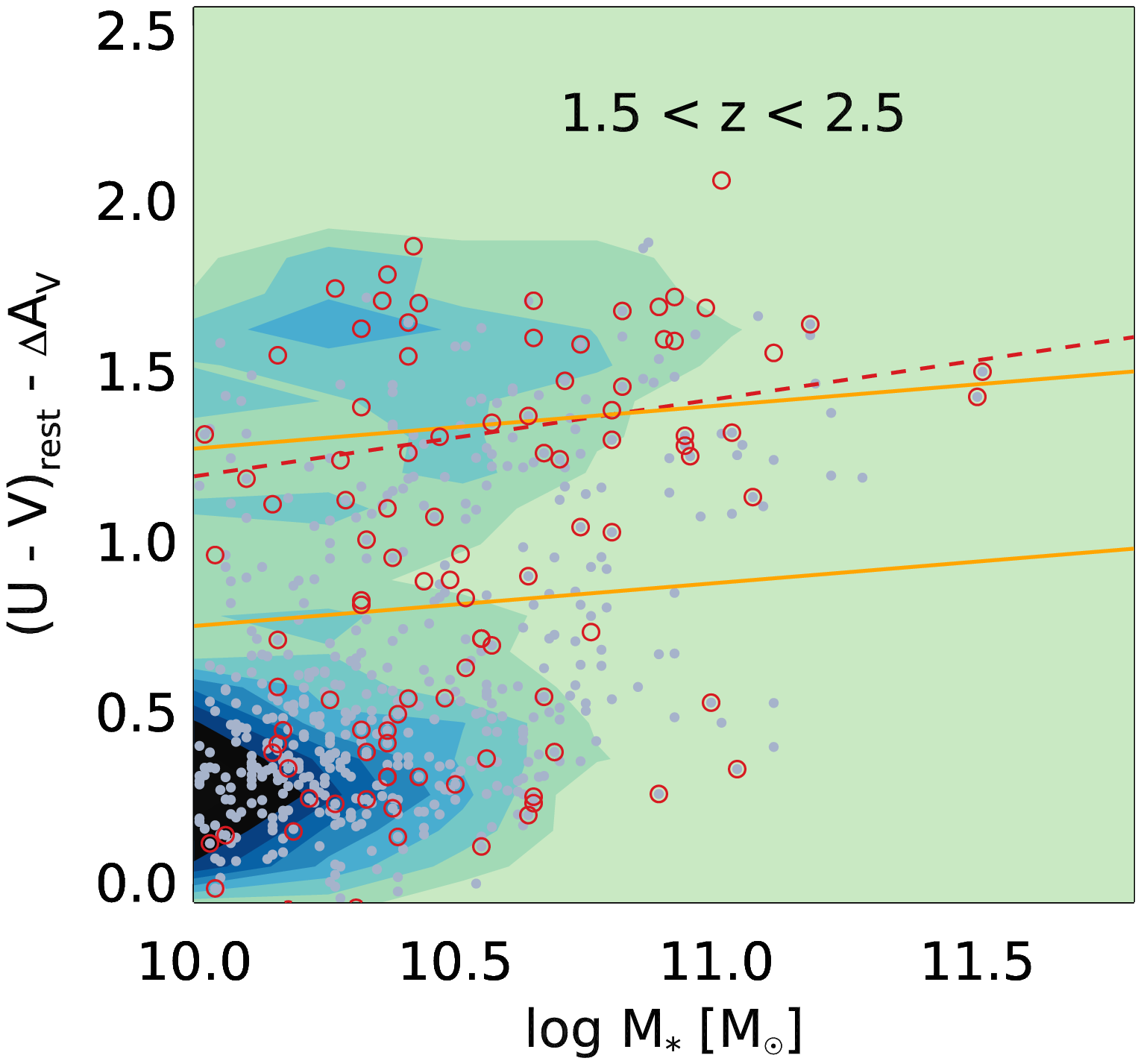}
\caption{Rest-frame color-mass relation (CMR), before (the upper two panels) and after (the lower two panels) extinction correction, for all galaxies in our sample. The left two panels show the CMR at $0.5 < z < 1.5$ while the right two panels show the CMR at $1.5 \leq z < 2.5$. AGN hosts and 24 $\mu$m-detected galaxies are denoted by red open circles and grey filled circles, respectively. The contours indicate the density distribution of galaxies in the CMR. The separations between red sequence and blue cloud based on observed colors (without extinction-correction) from \cite{Borch:2006} are shown by solid (top two panels) and dashed (bottom two panels) red lines. The orange lines in the bottom two panels indicate the separation between red sequence, green valley, and blue cloud based on extinction-corrected colors derived in this work.  After extinction correction, the fraction of red galaxies with 24$\mu$m detections ($f_{24\mu m} > 30 \mu$Jy) is significantly reduced (from $\sim$38\% to $\sim$14\% at $z \sim 1$ and $\sim$47\% to $\sim$17\% at $z\sim2$) suggesting that the extinction-corrected color successfully separates red galaxies due to dust attenuation from those due to old stellar populations.
\label{Fig:uv_lmass}}
\end{figure*}

\begin{figure*}[th!]
\begin{center}
\includegraphics[trim = 0 0 0 0, width=0.9\textwidth]{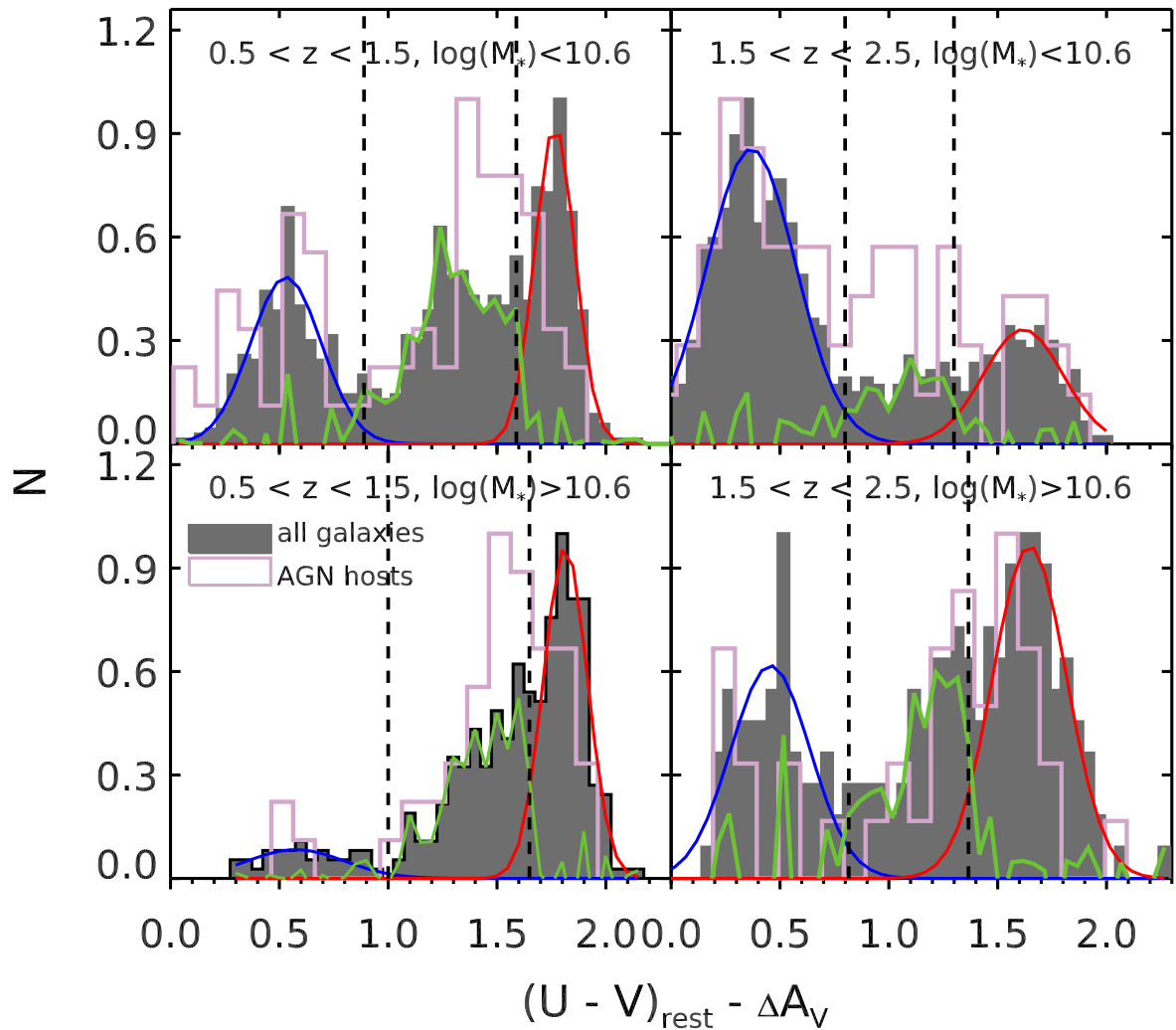}
\caption{Histograms of extinction-corrected rest-frame $U - V$ color for AGN hosts (solid magenta lines) and parent galaxies (gray filled histogram) in two ranges of stellar mass, $10 < log M_{\odot} < 10.6$ and $log M_{\odot} > 10.6$, respectively. The peak value of each histogram has been rescaled to unity. Red and blue lines are the Gaussian fitting results for red sequence and blue cloud, and the solid green lines mark the residual distribution after subtracting the sum of the two Gaussians. The vertical dashed lines indicate the boundaries of the green valley, where the residual distribution surpass the Gaussian fit for blue (blue-side) or red (red-side) galaxies.
\label{Fig:cmd_hist}}
\end{center}
\end{figure*}

\begin{figure*}[th!]
\begin{center}
\includegraphics[trim = -70 -60 80 70, angle=270,width=0.49\textwidth]{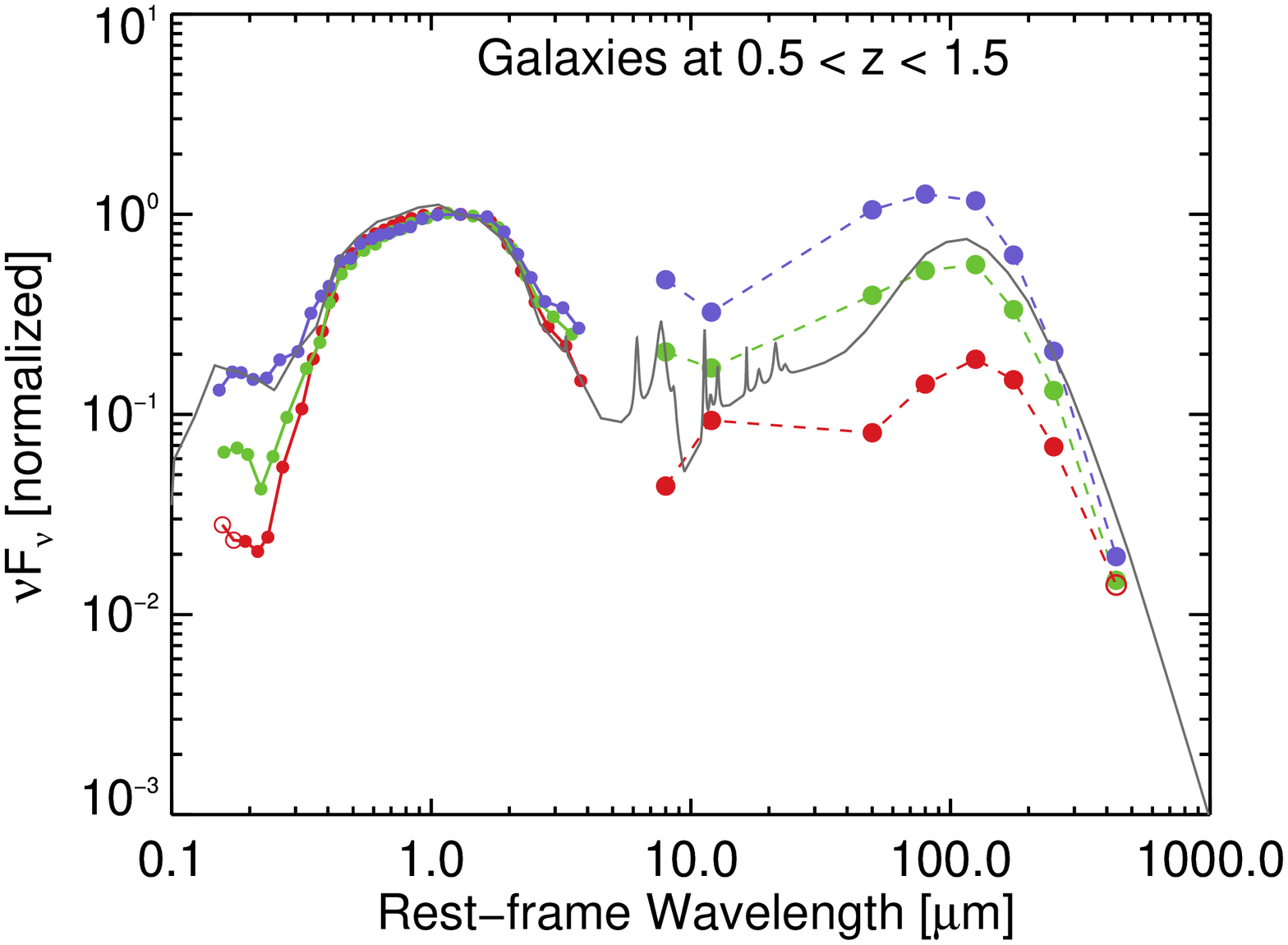}
\includegraphics[trim = -70 -60 80 70,angle=270,width=0.49\textwidth]{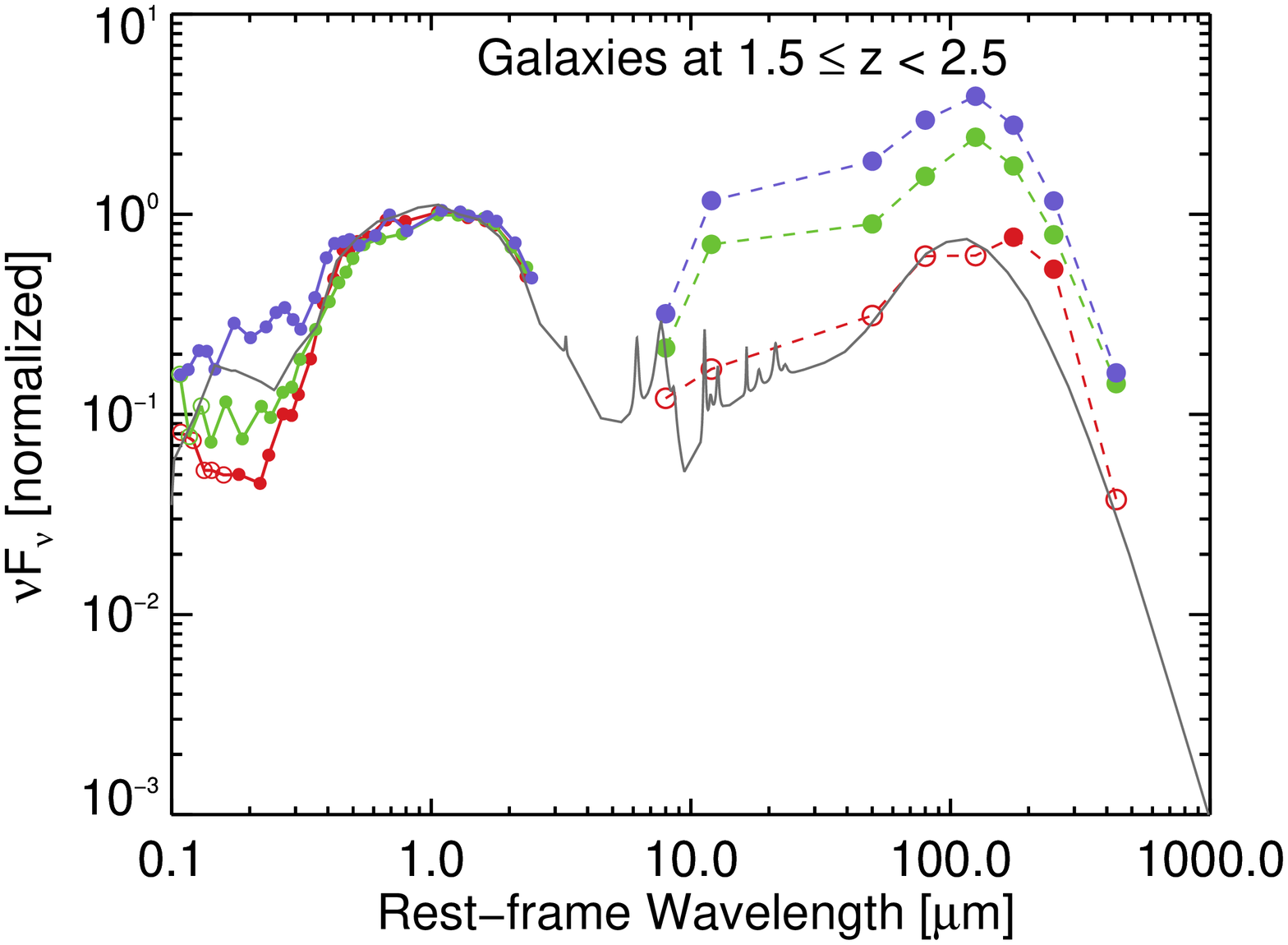}\\
%\vspace{0.2in}
\includegraphics[trim = -80 -60 40 70,angle=270,width=0.49\textwidth]{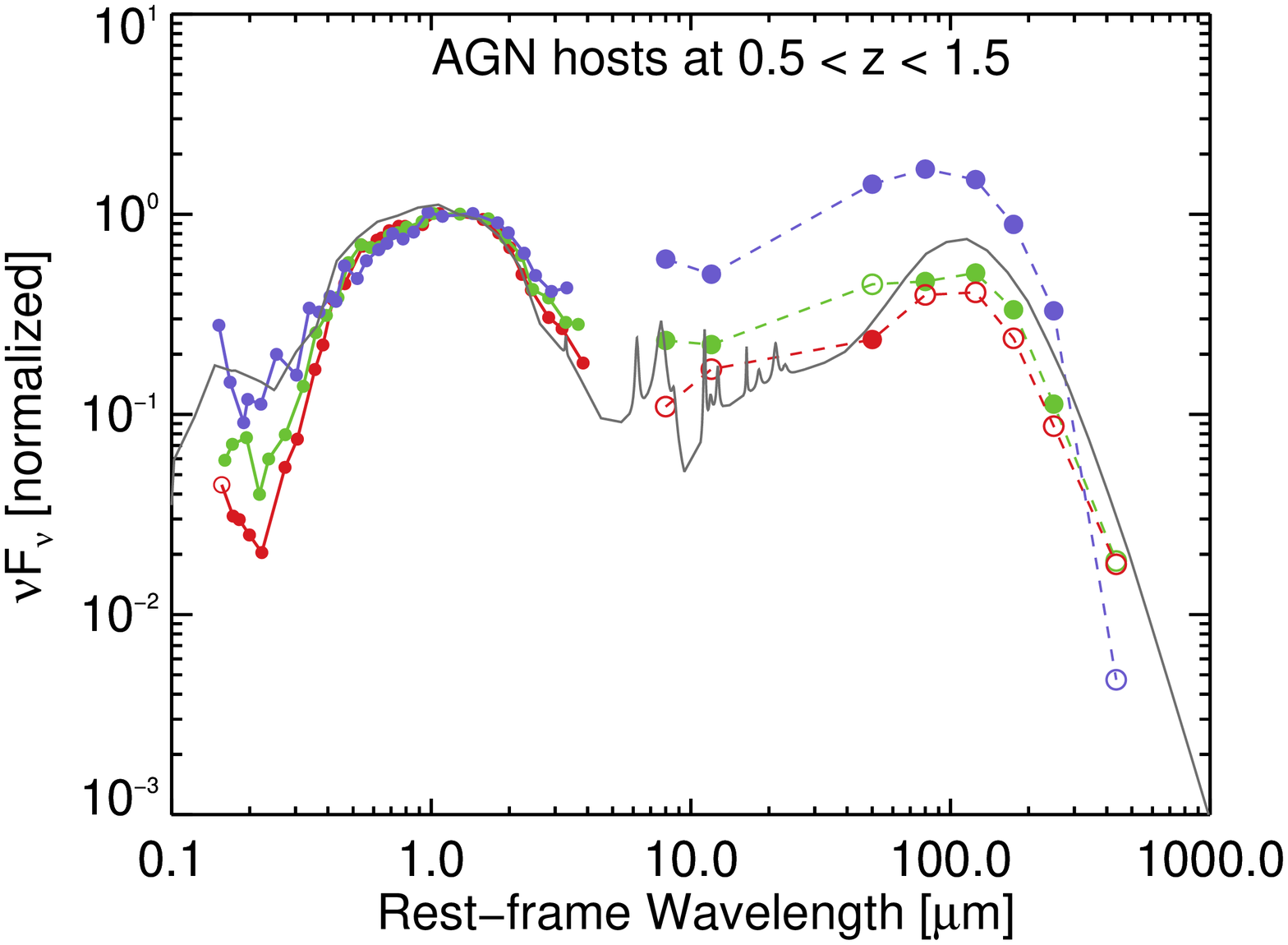}
\includegraphics[trim = -80 -60 40 70,angle=270,width=0.49\textwidth]{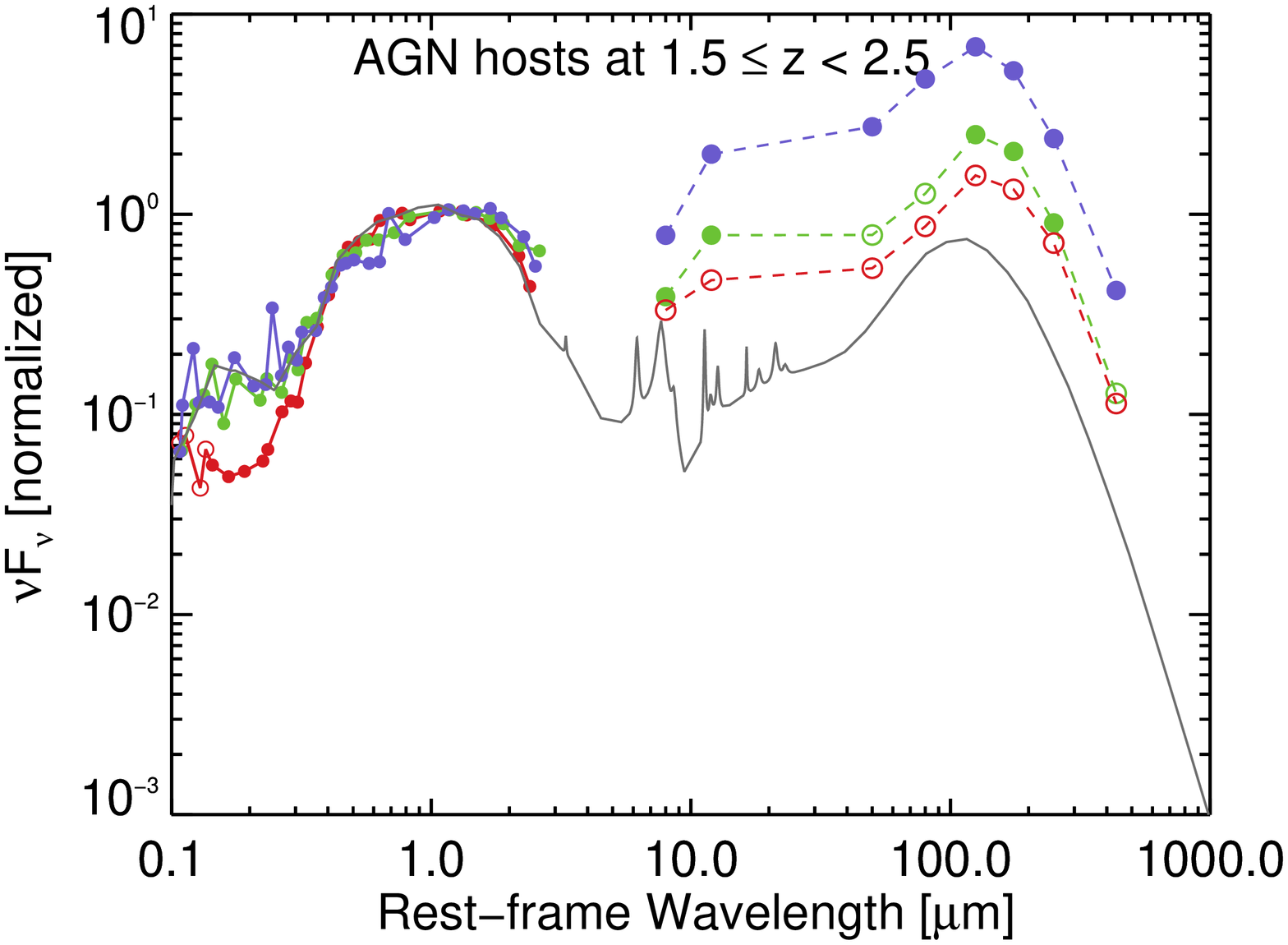}\\
\caption{Composite UV-to-FIR SEDs of the red, blue and green galaxies (the upper two panels) and AGN hosts (the lower two panels). The UV-to-NIR SEDs of galaxies in each population are de-redshifted to rest-frame and normalized to unity at J band ($\sim$1.25~$\um$), then projected onto a common wavelength grid. At each wavelength we derive the median and denote as filled symbols. If more than half of the sample is not detected within a wavelength grid, then we denote the median as open circles. The FIR fluxes from stacking are also normalized at J band. FIR data points for non-detections ($S/N < 3$) in the stacking are shown with their 3$\sigma$ upper limits, and are denoted by open circles. For reference, we over-plot a UV-to-FIR SED of a nearby Spiral galaxy (a blue galaxy based on our classification), M51, from the Grasil model \citep{Silva:1998}.
\label{Fig:median_sed}}
\end{center}
\end{figure*}

\begin{figure*}[th!]
\begin{center}
\includegraphics[trim = 0 0 0 0, width=0.49\textwidth]{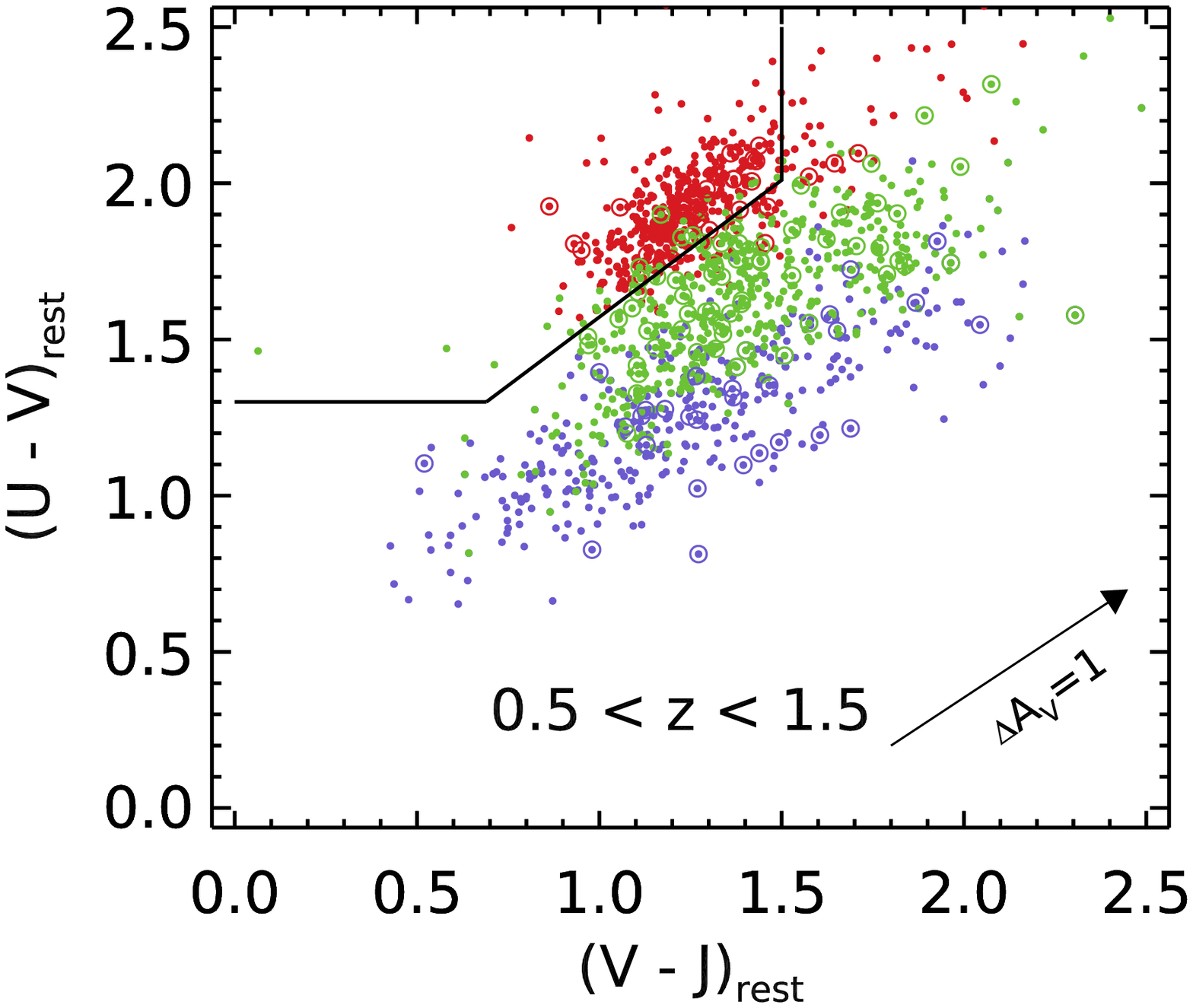}
\includegraphics[trim = 0 0 0 0,width=0.49\textwidth]{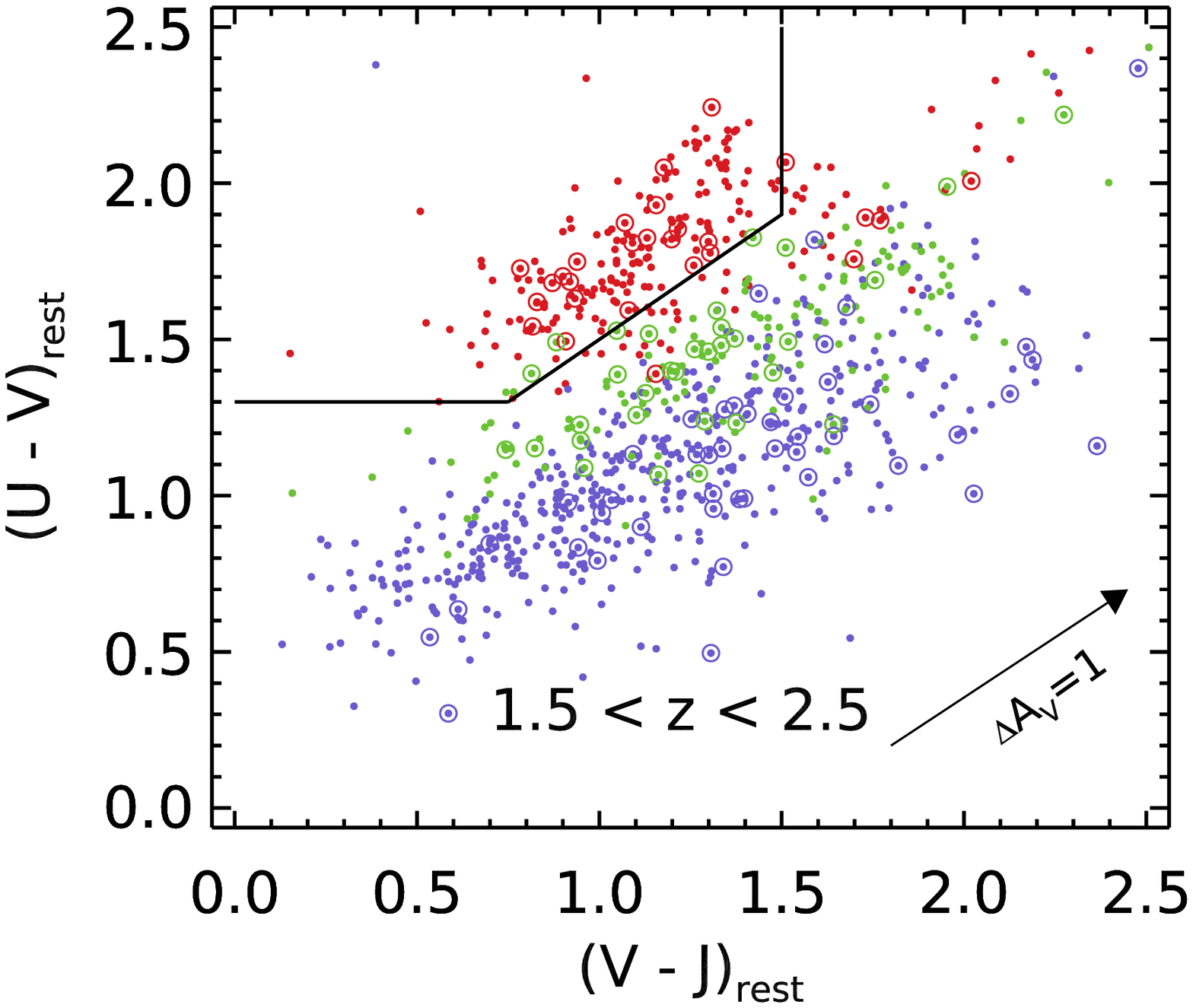}
\caption{Distribution of galaxies in rest-frame $U - V$ color vs. $V - J$ color. Red, blue, and green dots indicate red, blue and green galaxy populations, respectively, while the large circles represent AGN hosts in each population. The solid line denotes the dividing line between star-forming and quiescent galaxies~\citep{Whitaker:2012,Muzzin:2013b}. The distribution of our classified red and blue populations is consistent with 
expectations of quiescent and star-forming galaxies in this diagram. Moreover, the green populations fall in between them, consistent with being a transition population.
\label{Fig:uv_vj}}
\end{center}
\end{figure*}

The unprecedented multi-wavelength data set in the two fields allows us to select a complete massive galaxy sample with $M_{*} > 10^{10} M_{\odot}$
at $0.5 < z  < 2.5$ \citep{Grogin:2011}. Our final sample includes 2154 galaxies (1143 in GOODS-South and 1011 in GOODS-North) with $M_{*} > 10^{10} M_{\odot}$ and $0.5 < z < 2.5$, among which 221 (134 in GOODS-South and 87 in GOODS-North) host an AGN with 
$10^{41.9} < L_{X} < 10^{43.7}$ erg s$^{-1}$. The difference in the number of galaxies and AGNs in the two fields further implies that the major difference between the two fields is the depth of X-ray imaging. As a major consequence, most AGNs with $L_{X} < 10^{43}$ erg s$^{-1}$ and $z > 1.5$ are located in GOODS-South.%(Table~\ref{tab:sample_info}).

\section{Characterizing AGN hosts and non-AGN galaxies using extinction-corrected rest-frame colors}
In this section, we derive extinction-corrected rest-frame colors for all galaxies in our sample, and based on these colors, we classify our sample into red-sequence (red), green-valley (green), and blue-cloud (blue) galaxies. We then employ several methods to validate this classification.
\subsection{Extinction-corrected rest-frame colors}
We chose to use the extinction-corrected rest-frame $U - V$ colors to characterize galaxies in our sample because this color straddles the 4000~\AA~break in the galaxy spectrum and separates red galaxies dominated by older stellar populations from blue galaxies experiencing significant ongoing star formation. We have also confirmed that the basic results remain unchanged when using the extinction-corrected rest-frame $U - B$ color.

We first derived rest-frame $U - V$ for each galaxy with $EAZY$ by fixing its redshift to spectroscopic redshift if available or to our best-estimated photo-z.
Then we apply the reddening correction to the $U - V$ color and get the de-reddened color using $(U - V)^{\prime} = (U - V) - 0.47*A_{V}$, with 0.47 being the correction factor computed for the \cite{Calzetti:2000} extinction law \citep{Brammer:2009}. The dust extinction for each galaxy, $A_{V}$, is derived from SED fitting with FAST as mentioned in Section~\ref{Sec:properties}. \cite{Brammer:2009} have shown that with such derived $A_{V}$ (with the same stellar population synthesis model and dust extinction law as we use here), the extinction-corrected $U - V$ color successfully separated dusty star-forming galaxies from red and dead galaxies. We will further test this in Section~\ref{sec:validation}. In Figure~\ref{Fig:uv_lmass}, we plot the rest-frame $U - V$, both before and after de-reddening, versus stellar mass for AGN hosts and non-AGN galaxies in our sample. The sample is divided into two redshift bins, $0.5 < z < 1.5$ and $1.5 \leq z < 2.5$, which contain similar numbers of galaxies and AGN hosts.

As shown by the red lines in Figure~\ref{Fig:uv_lmass}, we have applied the (evolving) separation of ``red sequence'' from ``blue cloud'' galaxies following \cite{Bell:2004} and \cite{Borch:2006}, 
\begin{equation}
\label{Eqa:uv_nocor}
 (U - V)_\mathrm{rest} = 0.227 \mathrm{log_{10}}M_{*} -0.379 - 0.352z.\\ 
\end{equation}
A correction factor of 0.781 to the $U - V$ colors was applied to covert from Vega (which was used in \cite{Borch:2006}) to AB magnitudes. Though this dividing line is derived from galaxies at $z \lesssim 1$, it has been shown to be valid up to at least $z \sim 3$~\citep{Xue:2010}. However, since this dividing line is derived from observed colors (without extinction-correction), it remains unclear how it can be applied to extinction-corrected colors. To resolve this issue and search for a physically meaningful separation between different populations in the extinction-corrected color, we have performed a detailed analysis of the distribution of extinction-corrected colors in individual stellar mass and redshift bins. We divide both AGN hosts and parent galaxies into two stellar mass bins separated at $M_{*} = 10^{10.6} M_{\odot}$ to ensure that there are a similar number of AGNs in each bin.

Figure~\ref{Fig:cmd_hist} illustrates that the distributions of extinction-corrected $U-V$ colors in each stellar mass bin exhibit two apparent peaks, corresponding to the blue and red density peaks, respectively, in Figure~\ref{Fig:uv_lmass}, respectively. We then fit a single Gaussian to each peak. Based on the typical width of blue cloud and red sequence~\citep[see, e.g., ][]{JinS:2014}, we select the left 0.15 mag of red peaks as the red boundaries and the right 0.25 mag of blue peaks as blue boundaries during the fit, to reduce the effect of the overlapping region on the fitting results. We find that, after subtracting the sum of the two best-fit Gaussian models, substantial excess between the blue cloud and the red sequence is revealed, as shown by the solid green lines in Figure~\ref{Fig:cmd_hist}. We then define the intersection between this excess and blue cloud and red sequence as the boundaries of green valley, as shown by the vertical dashed lines.

Similar to the trend seen in Equation~\ref{Eqa:uv_nocor}, Figure~\ref{Fig:cmd_hist} reveals that the separation between red sequence and blue cloud (boundaries of green valley) in extinction-corrected $U - V$ also evolves with redshift and stellar mass, which tends to be redder with both increasing stellar mass and decreasing redshift.  Based on the constraints on the boundaries of green valley from the fit in the two stellar mass and redshift bins, we define the red and blue boundaries of green valley galaxies on the extinction-corrected color-mass diagram as:
\begin{align}
\label{Eqa:uv_cor}
 (U - V)_\mathrm{rest} - \Delta A_{V} = 0.126 \mathrm{log_{10}}M_{*} +0.58 - 0.286z\\
 (U - V)_\mathrm{rest} - \Delta A_{V} = 0.126 \mathrm{log_{10}}M_{*} - 0.24 - 0.136z
\end{align}
The slope to redder colors with increasing stellar mass and decreasing redshift flattens in Equation~\ref{Eqa:uv_cor} compared to Equation~\ref{Eqa:uv_nocor}, suggesting that part of the correlation shown in Equation~\ref{Eqa:uv_nocor} arises from the more severe dust extinction of galaxies at higher stellar mass and higher redshifts~\citep{Pannella:2015}. The different slopes with redshifts in the red and blue boundaries suggests a moderate broadening of the green valley with decreasing redshift, as can be seen in Figure~\ref{Fig:cmd_hist}, which changes from $\sim 0.52$ at $z \sim 2$ to $\sim 0.67$ at $z \sim 1$. We confirmed that slightly changing the definition, that is, applying $\pm$ 0.1 magnitudes shift on $U - V$, would not change significantly our results.

\subsection{Validation of the classification based on de-reddened colors}
\label{sec:validation}
To test whether the three galaxy populations indeed possess distinct properties, we construct composite rest-frame UV-to-FIR SEDs for different types of galaxies and AGN hosts, as shown in Figure~\ref{Fig:median_sed}. For the UV-to-NIR wavelengths we produce their median SEDs by first de-redshifting all galaxy photometry in the GOODS-North and GOODS-South catalogs to the  rest-frame using photometric redshifts described in Section~\ref{Sec:properties}, or spectroscopic redshifts when available. Then we normalize the individual SEDs to unity at rest-frame J band, that is, $\sim$ 12500 \AA, using our determined rest-frame $J$-band fluxes. Finally we project each SED on a common wavelength grid and derive the median SED. For the IR-to-submillimeter SEDs, we derive the fluxes at each wavelength by stacking across $Spitzer$ 16~\citep{Teplitz:2011} and 24 $\mu$m (GOODS-Spitzer, PI: M. Dickinson), $Herschel$ 100, 160, 250, 350 and 500 $\mu$m \citep{Elbaz:2011,Lutz:2011,Oliver:2012}, as well as  870 
$\mu$m imaging \citep{Borys:2003,Pope:2005,Hodge:2013}. Details of the stacking methods can be found in \cite{WangT:2016a,Schreiber:2015}.

\begin{figure*}[th!]
\includegraphics[trim=0 0 -100 0,angle=90,width=0.48\textwidth]{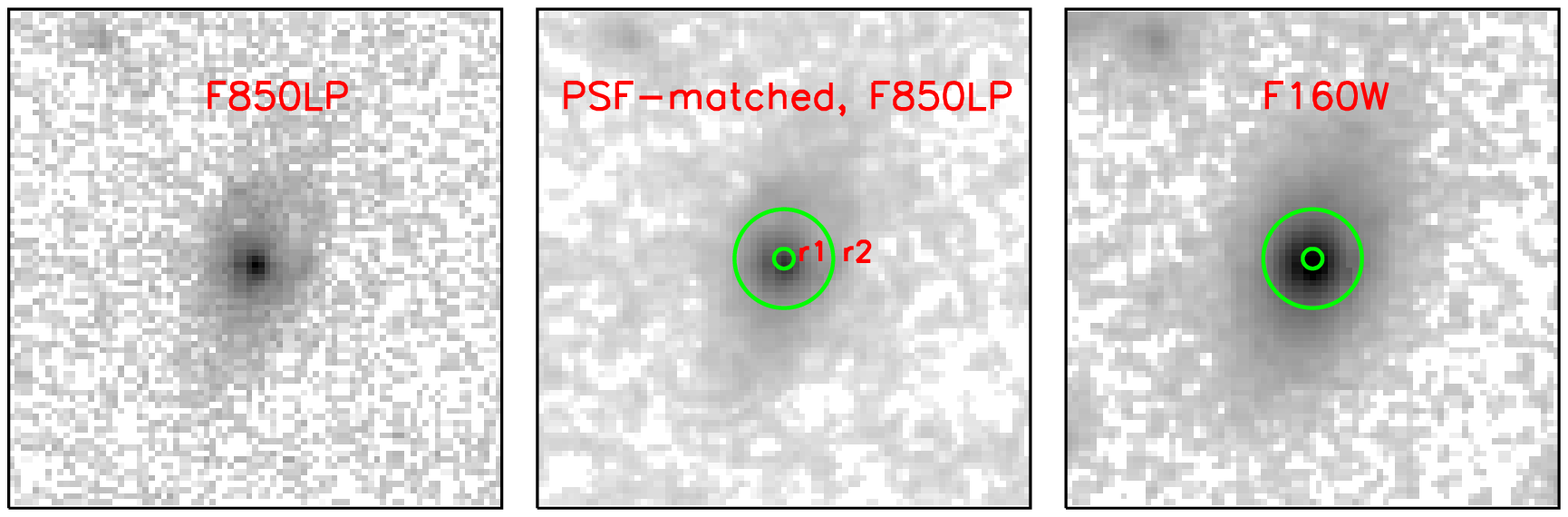}
\includegraphics[trim = 300 -150 40 70, angle=270,width=0.5\textwidth]{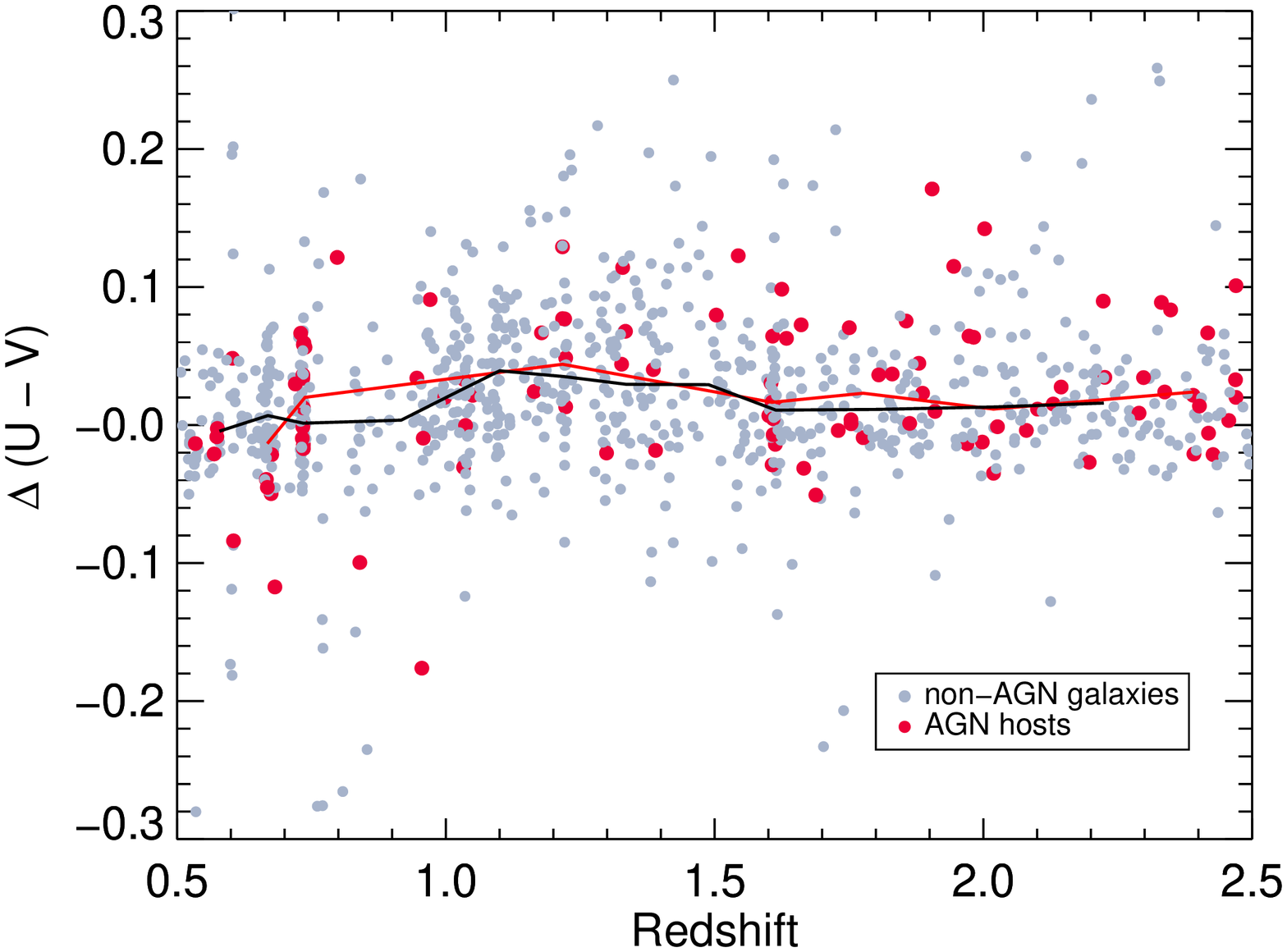}
\caption{Left: Representative sample images of one AGN in our sample in HST/ACS F850LP band, HST/ACS F850LP band (after PSF-matched) and WFC3/IR F160W images. The green circles represent the apertures we used to assess 
the AGN contribution on host colors. Right: redshift vs. $\Delta(U - V) $ for AGN hosts and non-AGN galaxies in GOODS-South. The red and black lines denote the sliding median for AGN hosts and non-AGN galaxies, respectively. \label{Fig:color_stamp}}
\end{figure*} 
We find that the red galaxies/AGN-hosts are much fainter in both UV and IR (including mid-IR and far-IR) than the blue ones, while the green ones lie in between. We also examine the median SEDs by separating galaxies and AGN hosts into different mass bins, which reveal similar trends. This illustrates that the three populations possess star formation activities consistent with their extinction-corrected colors.  Specifically for the PACS 100 $\mu$m band, we show the detection rates for the three populations, as shown in Table~\ref{tab:statistics}. At both $z \sim 1$ and $z \sim 2$, only very few ($<10\%$) of the red AGN hosts are significantly detected at 100 $\mu$m while the detection rates for the blue AGN hosts reach $\sim$70\% at $z \sim 1$ and $\sim$50\% at $ z\sim 2$.  Again, the detection rates of green AGN hosts lie in between, presenting further evidence that such classified galaxy populations have different levels of star-formation activity.

As a second test, we show the distribution of the three classes of galaxies in a rest-frame $U - V$ versus $V - J$ diagram, as shown in Figure~\ref{Fig:uv_vj}. 
In this diagram, star-forming galaxies with relatively unobscured star formation would have
both blue $U - V$ and $V - J$ colors.  On the other hand, both quiescent and dusty galaxies have red $U - V$ colors, yet dusty galaxies are significantly redder in $V - J$ than quiescent galaxies \citep{Williams:2009}. The red, green, and blue populations discussed here are in agreement with the UVJ classification with most red galaxies in the quiescent region, blue galaxies in the star-forming region, and the green galaxies in between. 

 \begin{figure*}[th!]
\begin{center}
\includegraphics[trim = 0 0 0 0, angle=0,width=\textwidth]{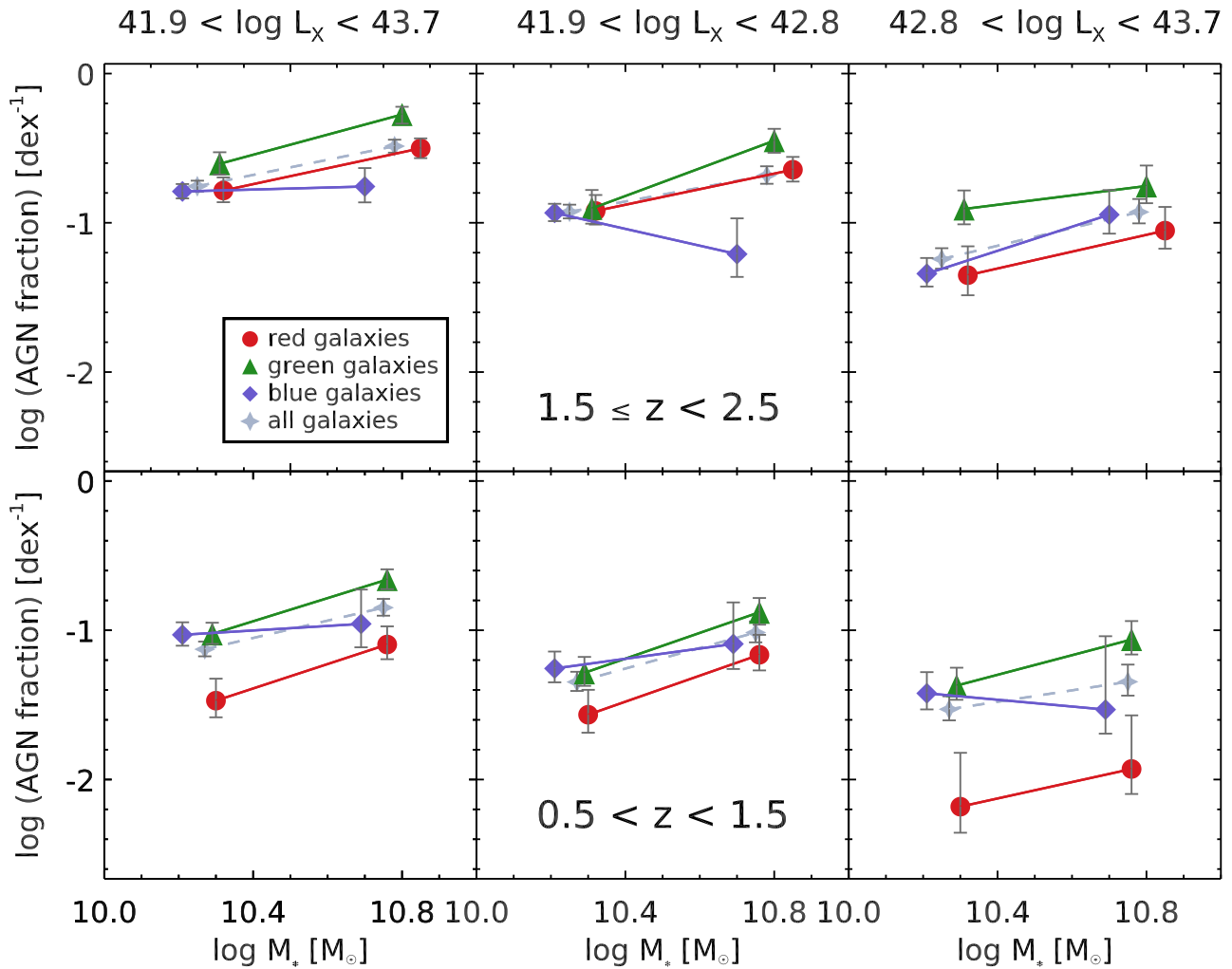}
\caption{The fraction of galaxies hosting an AGN as a function of stellar mass for different galaxy populations at $0.5 < z < 1.5$ (the bottom panels) and $1.5 \leq z < 2.5$ (top panels). We divide the sample into two stellar-mass bins separated at $10^{10.6} M_{\odot}$. The data points for each population are plotted at the median of each mass bin.
\label{Fig:p_lx_all}}
\end{center}
\end{figure*}

We conclude that, although the exact $A_{V}$ determination for individual galaxies may still suffer from the degeneracy between age and extinction~\citep[see, e.g., ][]{Smethurst:2015}, using the extinction-corrected $U - V$ color enables a statistically meaningful separation of galaxies with different star-formation properties. Moreover, the galaxies that we classify respectively as red, blue and green can be directly linked to the red-sequence, green-valley and blue-cloud galaxies in the local Universe from various SDSS studies, thus providing a convenient way to compare studies at low and high redshifts. 

\subsection{Assessing AGN contamination of host colors}
In this section, we evaluate whether or not host-galaxy colors are significantly contaminated by central AGNs.
Many previous studies have shown that in moderate 
luminosity AGNs, the AGN contribution to the total (galaxy+AGN) optical emission is small  \citep{Silverman:2008b,Xue:2010,Cardamone:2010,Simmons:2011}. Here, we further 
examine the AGN contamination of the determination of rest-frame colors. 
We first matched $HST$/ACS F606W ($V$ band) and F850LP ($z$ band) images to the same
resolution as that for WFC3 F160W ($H$ band) images through PSF matching, following the procedure described in \cite{Guo:2011}. Then, we measured aperture magnitudes in $V$, $z$, $H$ as well as F125W ($J$) band in two circular apertures centered at the
centroid of $H$-band emission with $r_{1}$ =  2 and $r_{2}$ = 10 pixels, corresponding to 0.12\arcsec and 0.6\arcsec, respectively. $(X - Y)$ colors, where $X$ and $Y$ is one of $VzJH$, were then measured with and without including the inner aperture, that is, $(X - Y)_{r_{2}}$ and $(X - Y)_{r_{2} - r_{1}}$. Since we did not remove stellar emission within the small 
aperture, the color difference $\Delta(X - Y) = (X - Y)_{r_{2}} - (X - Y)_{r_{2} - r_{1}}$ provides a upper limit on the color variations due to AGNs. 

For simplicity, we use $V - J$ and $z - H$ as a proxy of rest-frame $U - V$ colors for galaxies at $0.5 < z < 1.5$ and $1.5 \leq z < 2.5$, respectively. 
As shown in Figure~\ref{Fig:color_stamp}, we do not find significant difference in the distribution of $\Delta(U - V)$ between AGN hosts (with 10$^{41.9}$ erg s$^{-1}$ $< L_{X} <$ 10$^{43.7}$ erg s$^{-1}$) and non-AGN galaxies. The median color difference for AGN hosts and non-AGN galaxies is  $\sim$ 0.020 mag and $\sim 0.016$ mag, respectively.
Thus the integrated galaxy color does not appear to be significantly affected by the central AGN light, consistent with previous studies \citep{Pierce:2010b}.

\section{The dependence of AGN incidence on host color } 
In this section we derive the fraction of galaxies hosting an AGN  for the red, green, and blue populations, and examine whether or not there is an enhancement in the transition galaxies.

\subsection{AGN completeness weighting}
\label{subset:weighting}
Due to the varying depth of our X-ray data, both between the two fields and also within each field, we must correct for the incompleteness ( $V_{max}$ correction) in the data to get an unbiased view of the AGN population.

To do so, we use a method similar to \cite{Aird:2012}. For each galaxy population, we first split the sample into several redshift and mass bins. We denote the total number of galaxies in a bin as $N_{\rm gal}$. Then, within each bin, and for each X-ray source $i$ with measured X-ray luminosity $L_{X}^i$, we compute the number of galaxies ($N_{\rm gal}^i$) for which we could in principle detect an AGN of luminosity $L_{X}^i$, that is, galaxies that have $L_{X_{\rm limit}}^j \leq L_{X}^i$. The limiting luminosity $L_{X_{\rm limit}}^j$ is calculated using the redshift and X-ray sensitivity limits at its position, which is derived from the sensitivity map in \cite{Luo:2008} and \cite{Xue:2011} for GOODS-North and GOODS-South, respectively. We can then associate a weight $w_i$ to each detected AGN, with
\begin{equation}
\label{Lx_weight}
w_i = N_{\rm gal}/N_{\rm gal}^i\,.
\end{equation}
We then calculate the completeness-corrected AGN fraction in each mass and redshift bin using
\begin{equation}
{\rm AGN ~fraction} = \frac{1}{N_{\rm gal}} \sum_{i}^{N_X} w_i\,,
\end{equation}
where $N_X$ is the number of X-ray AGNs in the bin.
We also calculate the 68.3\% confidence intervals on the AGN fraction with Bayesian binomial statistics following \cite{Cameron:2011}. 
These completeness corrections make the reasonable assumption that the AGNs (and their host galaxies) that are not X-ray detected due to incompleteness have the same properties as the sources that are detected at the same \Lx, \Mstel, and $z$~\citep{Aird:2012}.

\subsection{The dependence of AGN fraction on stellar mass and X-ray luminosities for different hosts}
\label{subsec:f_agn}

To quantify the incidence of AGNs in different populations, 
we plot the fraction of galaxies hosting an AGN as a function of galaxy stellar mass, for red, blue, green as well as for the total galaxy population in Figure~\ref{Fig:p_lx_all}. Since the luminosity range of our X-ray AGN sample is relatively wide, we also show the AGN fraction separately for low- and high-luminosity AGNs separated at $L_{X} = 10^{42.8}$ erg s$^{-1}$ to explore the luminosity dependence of AGN fraction.

Figure~\ref{Fig:p_lx_all} reveals a general trend that the incidence of AGNs, that is, the AGN fraction, increases with increasing stellar mass, which is most pronounced for red and green galaxies. This trend with stellar mass appears to be weaker for blue galaxies. Furthermore, it also shows that the AGN fraction differs in different hosts at a given stellar mass. 
Most notably, red galaxies have the lowest probability of hosting an AGN at $z \sim 1$, particularly for high-luminosity AGNs (bottom right panel of Figure~\ref{Fig:p_lx_all}). Moreover, green galaxies show the highest  AGN fraction at both redshifts, which is most prominent at $z \sim 2$ and for the most massive galaxies with more than half of the green galaxies with $M_{*} > 10^{10.6} M_{\odot}$ hosting an X-ray AGN.
\subsection{The dependence of AGN fraction on redshift for different hosts}
\label{subsec:f_agn_z}

\begin{figure*}[!ht]
\begin{center}
\includegraphics[trim=0 0 0 0,angle=0,width=0.99\textwidth]{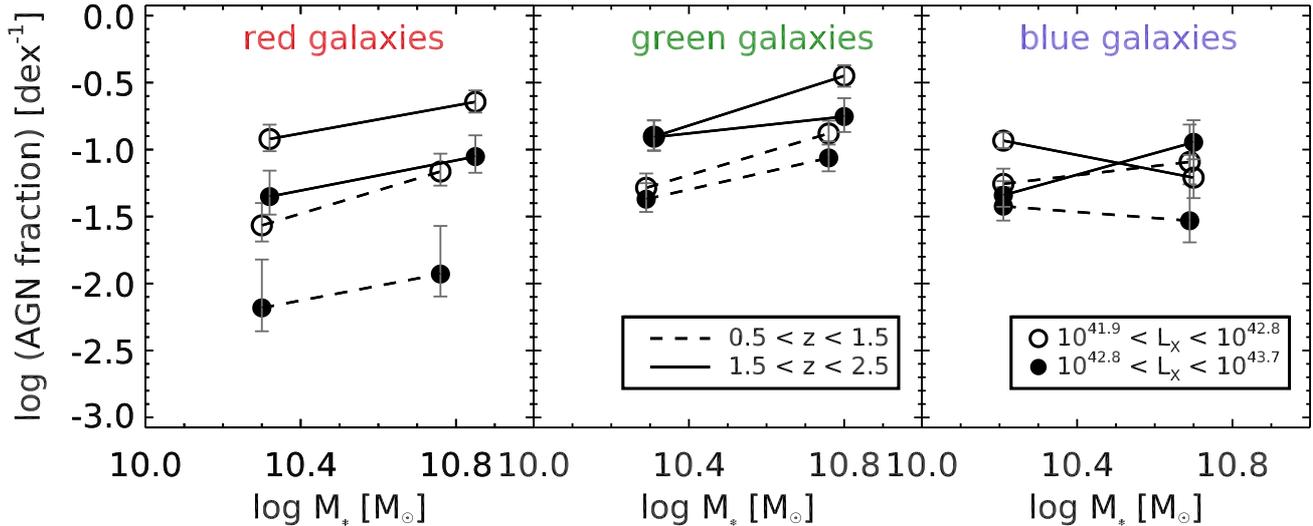}
\caption{As for Figure~\ref{Fig:p_lx_all}, except that we now show AGN fraction separately for each population. In each panel, the AGN fraction at $0.5 < z < 1.5$ and $1.5 \leq z < 2.5$ is shown with dashed and solid lines, respectively. While the AGN fraction in red galaxies increases by a factor of $\sim 5$ from $0.5 < z < 1.5$ to $1.5 \leq z < 2.5$, the AGN fraction in blue galaxies does not change much especially for massive galaxies with $M_{*} > 10^{10.6} M_{\odot}$.\label{Fig:p_lx_separate}}
\end{center}
\end{figure*}

To illustrate the evolution of AGN fraction with redshift more clearly, we re-plot the AGN fraction for each population separately in Figure~\ref{Fig:p_lx_separate}, which shows that the amplitude of the evolution of AGN fraction with redshift differs between different hosts. While the AGN fraction in red galaxies is higher by a factor of $\sim 5$ at $1.5 \leq z < 2.5$ (reaching $\sim 30\%$ at $M_{*} > 10^{10.6} M_{\odot}$) relative to $0.5 < z < 1.5$, the AGN fraction in blue galaxies does not change much with redshift, especially for massive galaxies with $M_{*} > 10^{10.6} M_{\odot}$, which remains basically the same at low and high redshifts. 

We conclude here that both the AGN fraction and its evolution with redshift are related to host color. This suggests that different hosts may have different modes of AGN accretion or the growth of SMBHs. Observationally, the mode of AGN accretion is directly reflected in the Eddington ratio distribution \citep{Kauffmann:2009,Trump:2011b,Aird:2012}. To reveal the mode of SMBH growth in different hosts and the physical mechanisms driving their growth, we thus proceed to study the Eddington ratio distribution for each population in the following section.

\section{The dependence of the Eddington ratio distribution on host colors}
\subsection{Determining the Eddington ratio distribution}
\label{subsec:Ledd}

\begin{table*}
\begin{center}
\caption{0.5-8 keV luminosities, Eddington ratios and host masses of AGNs in our sample.}
\begin{tabular}{lrrrrcccc}\hline\hline
Host galaxy & Number & Number of & Number &  Number of & AGN fraction & Median &  Median &  Median\\
color & of AGNs & FIR-AGNs\tablefootmark{a} & of galaxies & FIR-galaxies&  & log $L_{X}$  &  log $M_{*}$ &  log $\lambda_{\mathrm{Edd}}$\\
&  &  & &  & (\%) & (erg s$^{-1}$) &  ($M_{\sun}$) &   \\\hline
                          $0.5 < z < 1.5$ \\
Red          & 23	& 2 &   473  & 14&  5.1$^{+1.2}_{-0.8}$(4.9)\tablefootmark{b}	&   42.28	&       10.62  &        -2.09\\
Green 	  & 64 & 27	&   500  &182 & 13.2$^{+1.7}_{-1.4}$(12.8) 	&   42.67	&       10.60  &        -1.89\\
Blue	 & 28	& 20 &   325  & 182 &  9.5$^{+1.9}_{-1.4}$(8.6)      &   42.53      &       10.37  &          -1.57\\
All        & 115 & 49   & 1298 & 378  & 9.3$^{+0.9}_{-0.7}$(8.5)      &    42.54    &  10.51 &  -1.96\\
\hline
                         1.5 $\leq z < $2.5\\
Red	        & 27	& 1 &  226 & 6 &      22.5$^{+3.0}_{-2.5}$(11.9)	   &	42.86	&       10.67  &        -1.90\\
Green	 & 31	 & 8&  149 & 21 &      34.5$^{+4.1}_{-3.7}$(20.8)     &	42.99	&       10.49  &        -1.48\\
Blue		 & 47 & 24  &  481 & 101 &     16.4$^{+1.8}_{-1.6}$(9.8)       &	42.90     &       10.37  &        -1.41\\
All 	&  106 & 33    &   856 &128&      21.2$^{+1.5}_{-1.3}$(12.4)     & 42.90       &      10.44   &  -1.59\\
\hline
\end{tabular}
\tablefoot{
\tablefoottext{a}{The number of sources detected with $S/N>5$ at {\it Herschel}/PACS 100 $\um$.}\\
\tablefoottext{b}{The number quoted in parenthesis denotes the AGN fraction calculated without taking into account the varying depth of X-ray observations.}}
\label{tab:statistics}
\end{center}
\end{table*}
%.

\begin{figure*}[t!]
\begin{center}
\includegraphics[trim=30 0 0 -10]{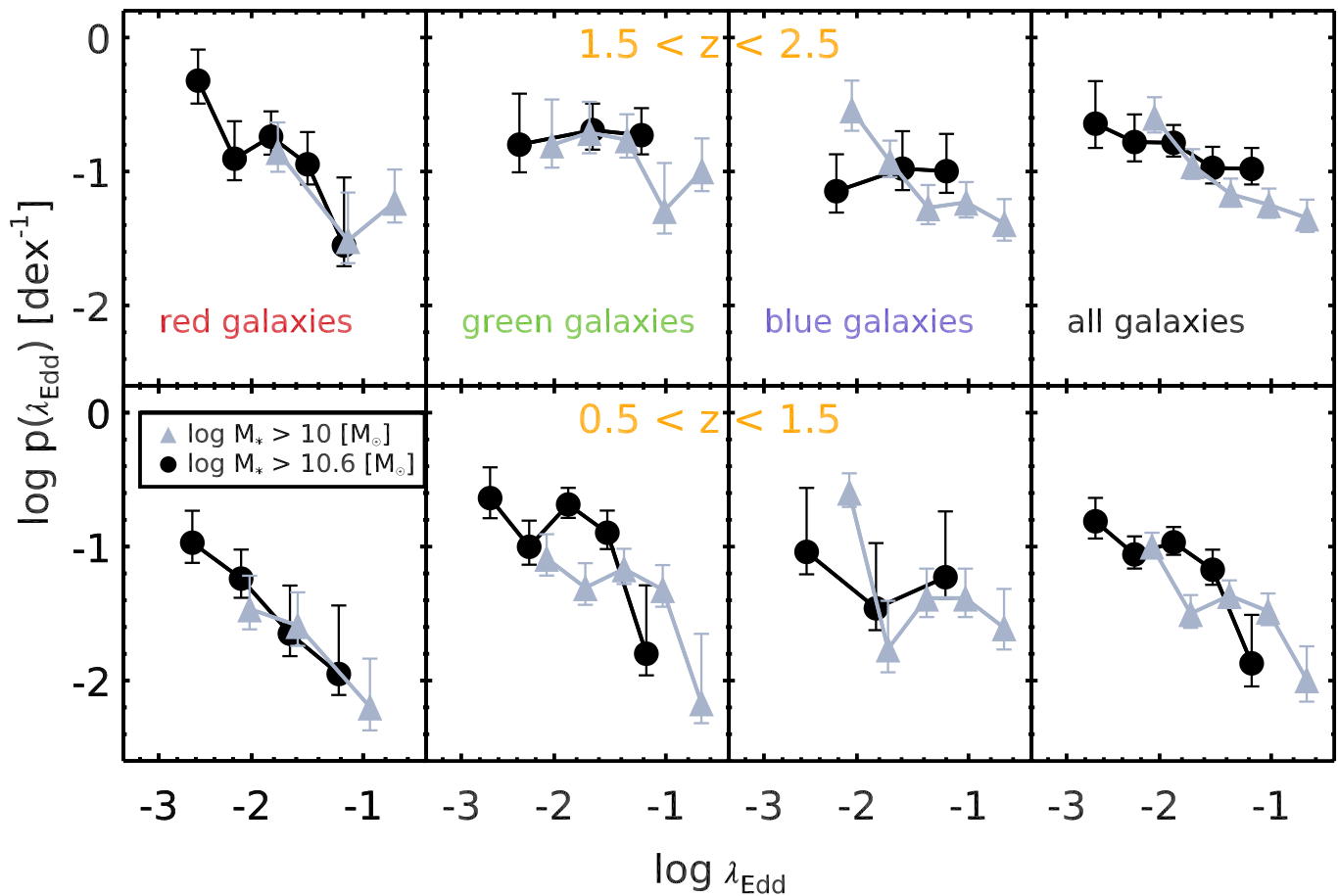}
\caption{The observed Eddington ratio distribution (after $1/V_{max}$ correction) for different galaxy populations, which are divided into two stellar-mass bins. This figure illustrates that the probability of hosting an AGN with certain $\lambda_{\mathrm{Edd}}$ does not have a strong dependence on stellar mass, at least across the stellar mass range 
of our sample. \label{Fig:ledd_dis_smass}}
\end{center}
\end{figure*}

\begin{figure*}[t!]
\begin{center}

\includegraphics[trim=30 0 0 0]{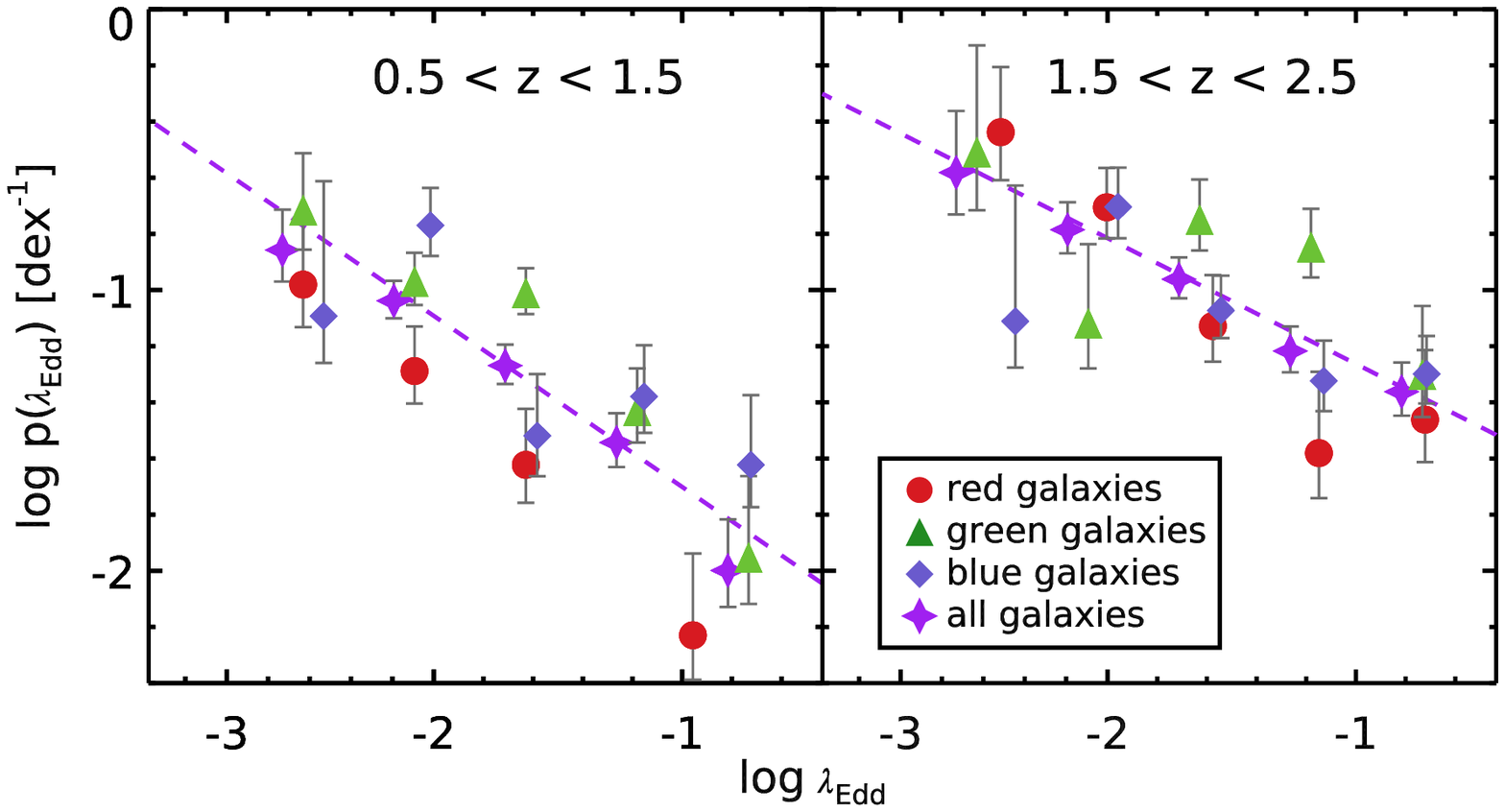}
\caption{As for Fig~\ref{Fig:ledd_dis_smass}, but here we show the $\lambda_{\mathrm{Edd}}$ distribution for the whole mass range of our sample.  
A linear fit of $\pledd$ for the total galaxy population is shown with the dashed purple line. This figure illustrates that the probability of hosting an AGN with certain $\lambda_{\mathrm{Edd}}$ is dependent on galaxy color, especially at lower redshift, that is, $z \sim 1$. \label{Fig:ledd_dis}}
\end{center}
\end{figure*}

The Eddington ratio, as defined by $\lambda_{\mathrm{Edd}} =   L_{\mathrm{bol}} / L_{\mathrm{Edd}}$, measures the specific accretion rates of the SMBH.
A measurement of $\lambda_{\mathrm{Edd}}$ requires measuring the bolometric luminosity of the AGN as well as its Eddington luminosity, which in turn depends on the mass of the SMBH. Direct measurements of SMBH masses rely upon the determination of the velocity dispersion of gas in the vicinity of the SMBH as provided by broad emission lines, and can only be performed on Type I AGNs with high-quality spectra. Alternatively, we can use the locally well-established $M_{BH}-M_{Bulge}$ relation to estimate SMBH masses, assuming that this relation has no strong evolution since $z \sim 2.5$. However, we do note that there is still much debate over whether or not and how the $M_{BH}-M_{Bulge}$ relation evolves with redshift \citep{Peng:2006,Merloni:2010,ShenY:2010}. On the other hand, \cite{Jahnke:2009} suggest that while the $M_{BH}-M_{Bulge}$ relation may evolve with redshift, the correlation between $M_{\mathrm{BH}}$ and the total stellar mass remains the same as the local  $M_{BH}-M_{Bulge}$ relation up to $z \sim 2$ \citep[also see, e.g., ][]{Schramm:2013}. Therefore, here we applied the local $M_{\mathrm{BH}}-M_{\mathrm{Bulge}}$ relation to the total stellar mass of galaxies in our sample to derive the black-hole mass. In this way despite the uncertainties in estimating the black hole mass, we can consider the  $\lambda_{\mathrm{Edd}}$ as a tracer of the specific accretion rates of galaxies, that is, the rate of black hole growth relative to the stellar mass of the host galaxy \citep[see, e.g., ][]{Aird:2012}.

Assuming that the AGN hosts follow the local relation between SMBH mass and host mass for spheroidal galaxies $M_{\rm BH}$= $\mu M_{*}$ (where $\mu \approx$ 0.0014, \citealt{Haring:2004}) and we can estimate $\lambda_{\mathrm{Edd}}$ for the AGNs as 
\begin{equation}
\label{equation:Ledd}
\lambda_{\mathrm{Edd}} =  \frac{ L_{bol}} {L_{Edd}} = \frac{\eta L_X}{1.26 \times 10^{38}(\frac{M_{\rm BH}}{  M_\odot})} = \frac{\eta L_X}{1.26 \times 10^{38}(\frac{\mu M_{*}}{  M_\odot})}
\end{equation}
where $\eta $ is the bolometric correction of the X-ray emission at 0.5$-$8 keV, and $L_{X}$ is in units of erg s$^{-1}$. We use a constant  bolometric correction factor of  $\eta$ = 20, which is typical for local $L_{X} = 10^{42-44}$ ergs s$^{-1}$ AGNs \citep{Marconi:2004,Vasudevan:2007}.

For each population, we then calculate the fraction (or probability) of galaxies in a given mass and redshift bin of hosting an AGN with Eddington ratio $\lambda_{\mathrm{Edd}}$, and denote it as $p(\lambda_{\mathrm{Edd}} \giv \mstel,z)$ ($p(\lambda_{\mathrm{Edd}})$, hereafter). Following the way we correct for the incompleteness in AGN fraction in Section \ref{subset:weighting} (essentially a 1/$V_{max}$ correction), for each bin of $\lambda_{\rm Edd}$, each AGN $i$ of $\lambda_{\rm Edd} = \lambda_{\rm Edd}^i$ falling in that bin is weighted by $w_{i} = N_{\rm gal}/N_{\rm gal}^i$, in which $N_{\rm gal}$ denotes the total number of galaxies in the redshift and mass bin while $N_{\rm gal}^i$ denotes the number of galaxies (among $N_{\rm gal}$) with $\lambda_{Edd_{\rm limit}}  \leq \lambda_{\rm Edd}^i$. $\lambda_{Edd_{\rm limit}}$ is determined by the redshift of each galaxy and the X-ray depth at their positions. Then the Eddington ratio distribution for galaxies at each redshift and stellar mass bin can be denoted as:
\begin{equation}
\label{equ:pledd_def}
p(\lambda_{\mathrm{Edd}} \giv \mstel,z) = \sum_{i}^{N_X} w_i/N_{gal}\
\end{equation}
 with $N_{X}$ being the number of AGNs falling in each bin (in logarithm) of $\lambda_{\mathrm{Edd}}$. The binning of $\lambda_{\mathrm{Edd}}$ is based on the expected range of $\lambda_{\mathrm{Edd}}$ in each parent galaxy sample given the stellar mass and X-ray luminosity range. $\pledd$ is then normalized by the size of a bin in log $\lambda_{\mathrm{Edd}}$.

\subsection{Mass dependence of Eddington ratio distribution for galaxies with different colors}
\label{subSec:ledd_mass}
We first examine whether or not there is strong mass dependence of $\pledd$ for different populations. Figure~\ref{Fig:ledd_dis_smass} presents $\pledd$ for each population and also the total galaxy population divided into two stellar mass bins, which 
shows that in almost all bins of $\ledd$, $\pledd$ is in good agreement (within 1-2$\sigma$) between the low-mass and high-mass samples. For the total galaxy sample, which has a relatively large number of AGNs, we perform a simple linear fit for the low-mass and high-mass samples, which is defined as:
\begin{equation}
\label{equ:pledd_noz}
p(\lambda_{\mathrm{Edd}})~d\log \lambda_{\mathrm{Edd}}  = A\left(\frac{\lambda_{\mathrm{Edd}} }{\lambda_{\mathrm{Edd_{cut}}}}\right)^{-\alpha}~d\log \lambda_{\mathrm{Edd}}. 
\end{equation}
$\lambda_{\mathrm{Edd_{cut}}}$ is an arbitrary scaling factor, which we adopt at the Eddington luminosity, that is, $\lambda_{\mathrm{Edd_{cut}}} = 1$. This power-law distribution of $\pledd$ is also suggested in several other previous studies at lower redshifts~\citep{Aird:2012,Trump:2015,Jones:2016}. 
The best-fit power-law slopes are in agreement with each other at 1 $\sigma$ level. This suggests that the mass dependance of $\pledd$ for all the galaxy populations across $0.5 < z < 2.5$ is relatively weak, if there is any, at least for the stellar mass range of our sample ($M_{*} > 10^{10} M_{\odot}$). This confirms previous arguments for Type-II AGNs at $z \sim 0$~\citep{Kauffmann:2009} and X-ray AGNs at $z \sim 0.6$~\citep{Aird:2012}, and extends to much higher redshifts. Though in a few cases, a mild dependence on stellar mass (in terms of both normalization and shape) cannot be fully ruled out, for example, blue galaxies  at $1.5 < z < 2.5$,  the limited statistics of our data and relatively narrow stellar mass range inhibits further quantitative constraints on this issue.

Considering the absence of a strong mass dependence of $\pledd$ for all the three galaxy populations, we re-calculate $\pledd$ without dividing into different mass bins, as shown in Figure~\ref{Fig:ledd_dis}.
Figure~\ref{Fig:ledd_dis} reveals immediately that the normalization of $p(\lambda_{\mathrm{Edd}})$ is not the same for different hosts, which is consistent with the distinct AGN fraction we observed in Section~\ref{subsec:f_agn}. At $z \sim 1$, the red galaxies show significantly lower probability of hosting AGNs at nearly all $\lambda_{\mathrm{Edd}}$ than green or blue galaxies. At $z\sim 2$, these differences tend to diminish except that a higher probability of hosting AGNs with $\lambda_{\mathrm{Edd}} \gtrsim 10^{-2}$ is revealed in green galaxies. We perform the same power-law fit to different populations and list their best-fit parameters and corresponding (reduced) $\chi^2$ values in Table~\ref{tab:linefit}.

Based on the best-fit $\chi^2$ values of the linear fit of $\pledd$, we find that $\pledd$ for all three galaxy populations, as well as the total galaxy, is fully consistent with a power-law distribution, as described in Equation~\ref{equ:pledd_noz}. In particular, $\pledd$ for red and total galaxy populations can be perfectly described by the power-law distribution at both redshifts, with a power law slope of $\sim 0.6$ and $\sim 0.4-0.5$, respectively.
On the other hand, the relatively poor power-law fit of $\pledd$ for blue and green galaxies mostly originates from a lack of low-$\ledd$ in blue galaxies and an enhancement of high-$\ledd$ AGNs in green galaxies, respectively. However, we note that these fitting results (particularly when separated into different populations) suffer from uncertainties from small number statistics, as well as density fluctuations in each bin. These effects are most pronounced in the lowest and highest $\ledd$ regime.

\begin{table*}
\begin{center}
\caption{Best-fit parameters from linear fitting of the Eddinton ratio distribution for different hosts\label{tab:linefit}}
\begin{tabular}{cccccccc}\hline\hline
& Red galaxy    &  Green galaxy   & Blue galaxy & All galaxies & \\
 \multicolumn{1}{l}{$0.5 < z < 1.5$}  &\\
\cline{1-1}

$log A$          & -2.82$\pm0.35$   &      -2.07$\pm0.22$ & -2.01$\pm0.23$  & -2.21$\pm0.17$\\
$\alpha$     & 0.66$\pm0.16$      &      0.52$\pm0.11$    & 0.48$\pm0.13$    & 0.51$\pm0.09$\\
$\chi^2$(reduced)     & 0.04 (2\tablefootmark{a})             &      1.20 (3)            & 1.99 (3)            & 0.44 (3)\\

\multicolumn{1}{l}{$1.5 < z < 2.5$} &\\
\cline{1-1}
$log A$           &     -2.04$\pm0.25$    & -1.23$\pm0.22$ & -1.61$\pm0.17$  & -1.64$\pm0.13$\\
$\alpha$      &     0.58$\pm0.13$     & 0.22$\pm0.13$  & 0.35$\pm0.11$    & 0.37$\pm0.08$ \\
$\chi^2$(reduced)        &     0.65 (3)            & 1.92 (3)           & 1.40 (3)           & 0.07 (3) \\
\hline

\end{tabular}
\tablefoot{
\tablefoottext{a}{Degrees of freedom.}}
\end{center}

\end{table*}

\begin{figure*}[t!]
\begin{center}
\includegraphics[width=0.8\textwidth]{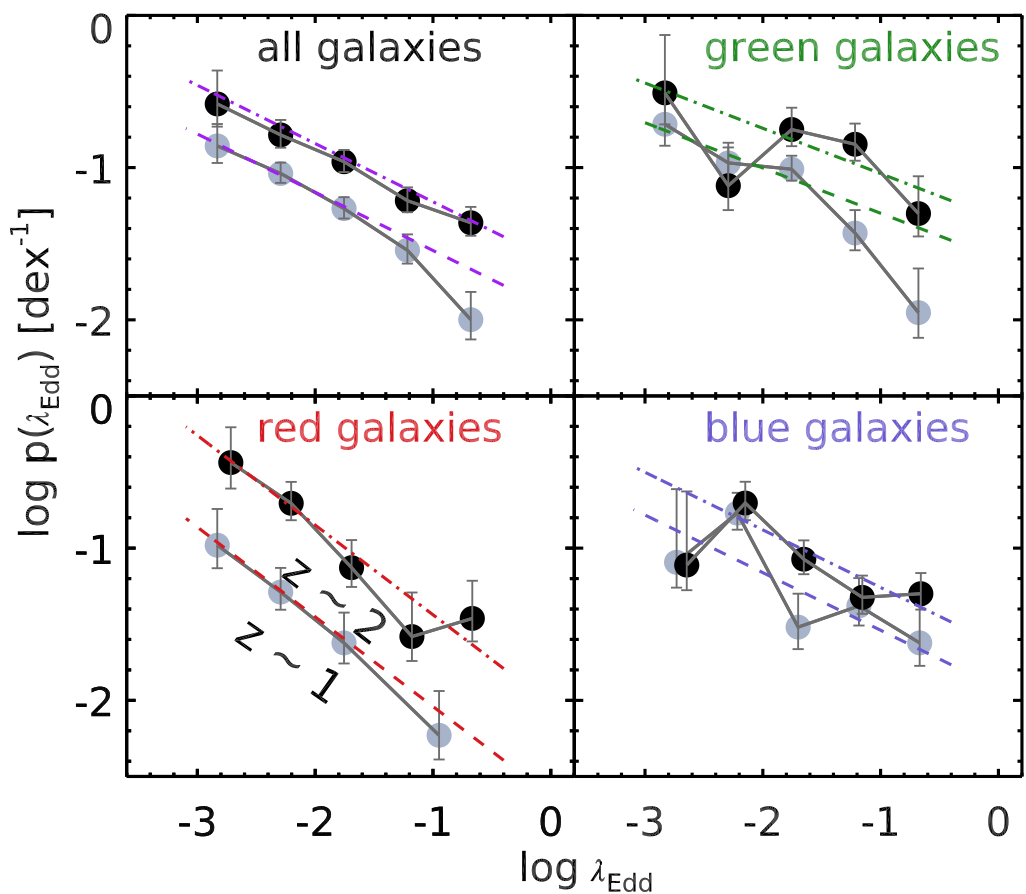}
\caption{As for Fig~\ref{Fig:ledd_dis}, but here we show the binned $\lambda_{\mathrm{Edd}}$ distribution for each population separately. In each panel, the $\lambda_{\mathrm{Edd}}$ distribution at $z \sim 2$ and $z \sim 1$ is shown with filled squares and circles, respectively. The best-fit models of $\pledd$ from MXL fitting of the unbinned data (Table~\ref{tab:MXLfit}) at two redshifts ($z \sim 1, 2$) are shown by dashed and dot-dashed lines, respectively.
%The grey lines show the observed (binned) $\lambda_{\mathrm{Edd}}$ distribution of local Type II AGN hosts with SMBH mass of $10^{7} <  M_{BH} < 10^{8} M_{\odot}$ with different $D_{n}$(4000) index from \cite{Kauffmann:2009}.  
\label{Fig:ledd_dis_z}}
\end{center}
\end{figure*}

 \begin{figure}[]
\begin{center}
\includegraphics[width=0.5\textwidth]{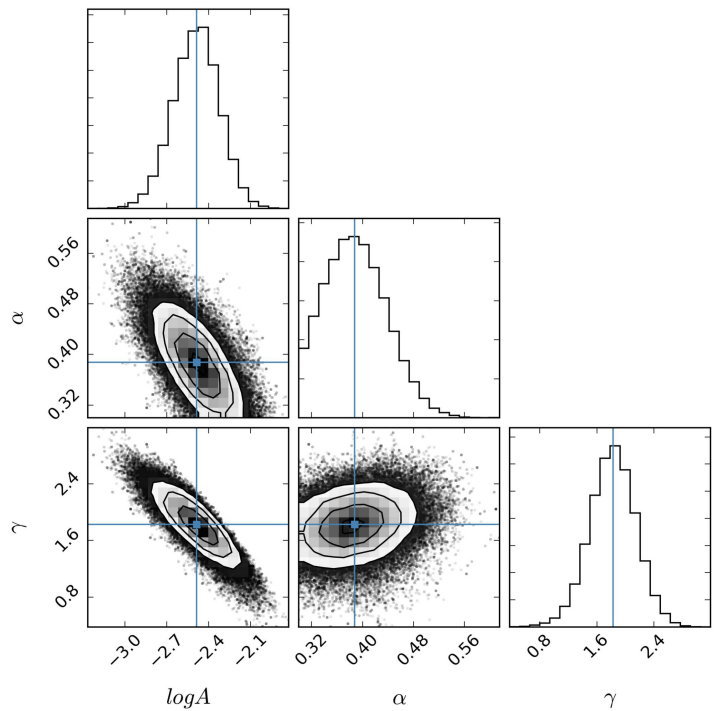}
\caption{Posterior probability distributions of the three parameters defining $\pledd$ in Equation~\ref{equ:pledd_noz} for the whole galaxy population (all galaxies) across $0.5 < z < 2.5$ from the MCMC analysis. 
The solid lines denote the best-fit values for each parameter, which are also listed in Table~\ref{tab:MXLfit}. The density contours show the 68.3, 90, and 95 percent posterior confidence probabilities. \label{Fig:MCMC_corner}}
\end{center}
\end{figure}

\begin{table*}
\begin{center}
\caption{Best-fit parameters from maximum-likelihood fitting of Eddington ratio distribution with redshift evolution\label{tab:MXLfit}}
\begin{tabular}{llllllllll}\hline\hline
&  Red galaxies    &  Green galaxies   & Blue galaxies & All galaxies & \\
\hline
\multicolumn{1}{l}{$0.5 < z < 1.5$}  &\\
\cline{1-1}
$logA$       & -3.3$^{+0.5}_{-0.5}$   &      -2.0$^{+0.2}_{-0.3}$    & -2.3$^{+0.3}_{-0.3}$   & -2.2$^{+0.2}_{-0.2}$\\
$\alpha$     & 0.57$^{+0.13}_{-0.15}$ &      0.37$^{+0.10}_{-0.10}$  & 0.44$^{+0.15}_{-0.15}$ & 0.40$^{+0.07}_{-0.07}$\\
$\gamma$     & 2.3$^{+1.4}_{-1.3}$    &      0.6$^{+0.6}_{-0.4}$     & 0.8$^{+0.7}_{-0.6}$    & 0.7$^{+0.6}_{-0.4}$ \\
\multicolumn{1}{l}{$1.5 < z < 2.5$}  &\\
\cline{1-1}
$logA$       & -3.1$^{+0.7}_{-0.7}$   &      -1.6$^{+0.3}_{-0.3}$    & -3.1$^{+0.7}_{-0.6}$   & -2.8$^{+0.5}_{-0.5}$\\
$\alpha$     & 0.56$^{+0.14}_{-0.15}$ &      0.20$^{+0.14}_{-0.12}$  & 0.36$^{+0.12}_{-0.12}$ & 0.38$^{+0.08}_{-0.08}$\\
$\gamma$     & 2.5$^{+1.5}_{-1.4}$    &      1.1$^{+0.6}_{-0.6}$     & 3.2$^{+1.1}_{-1.4}$    & 2.5$^{+1.0}_{-1.0}$ \\
\multicolumn{1}{l}{$0.5 < z < 2.5$}  &\\
\cline{1-1}
$logA$       & -3.7$^{+0.4}_{-0.4}$   &      -2.0$^{+0.2}_{-0.3}$    & -2.4$^{+0.3}_{-0.3}$   & -2.5$^{+0.2}_{-0.2}$\\
$\alpha$     & 0.59$^{+0.11}_{-0.12}$ &      0.30$^{+0.08}_{-0.08}$  & 0.38$^{+0.08}_{-0.09}$ & 0.38$^{+0.05}_{-0.05}$\\
$\gamma$     & 3.4$^{+0.7}_{-0.7}$    &      1.5$^{+0.5}_{-0.5}$     & 1.6$^{+0.6}_{-0.6}$    & 1.8$^{+0.3}_{-0.3}$ \\

\hline

\end{tabular}
\end{center}
\end{table*}

\subsection{Redshift evolution of the Eddington ratio distribution in different host galaxies}
\label{subsec:ledd_z}
We further plot the $\lambda_{\mathrm{Edd}}$ distribution for each population separately in Figure~\ref{Fig:ledd_dis_z} to study their evolution with redshift. In each panel, we plot $p (\lambda_{\mathrm{Edd}})$ at two redshifts,  $z \sim 1$ and $z \sim 2$. 
A general trend is immediately seen whereby the probability of hosting an AGN with a given $\lambda_{\mathrm{Edd}}$ increases with redshift in all types of galaxy. However, the rate of the increase turns out to be strongly dependent on host color. Among the three populations, red galaxies show the strongest redshift evolution with $p (\lambda_{\mathrm{Edd}})$ being approximately five times higher at $z \sim 2$ than $z \sim 1$. On the other hand, $p (\lambda_{\mathrm{Edd}})$ for blue galaxies only increases by a factor of approximately 1.4 from redshift $z \sim 1$ to $z \sim 2$. For green galaxies, a strong evolution at $\lambda_{\mathrm{Edd}} \gtrsim 10^{-2}$ is revealed, which is three times higher at $z \sim 2$ than at $z \sim 1$.

To put direct constraints on the form of the observed redshift evolution, we incorporated a simple power-law form of redshift evolution into the expression of 
$\pledd$ in Equation~\ref{equ:pledd_noz} as : 
\begin{equation}\label{equ:pledd_withz}
p(\lambda_{\mathrm{Edd}})~d\log \lambda_{\mathrm{Edd}} = A\left(\frac{\lambda_{\mathrm{Edd}} }{\lambda_{\mathrm{Edd}_{cut}}}\right)^{-\alpha}(1+z)^{\gamma}~d\log \lambda_{\mathrm{Edd}}.
\end{equation} 
To simultaneously fit the normalization, power-law slope, and the factor of redshift evolution, we employ an Extended Maximum Likelihood (MXL) fit of unbinned $\ledd$ data~\citep[see, e.g.,][] {Aird:2012}. The MXL fit removes redshift and $\ledd$-binning and hence avoids the effect of density fluctuations among different bins on the overall distribution. It also uses all the information of individual galaxies in the sample, while the 1/$V_{max}$ approach loses information on the locations of AGNs in each $\ledd$ bin. These permit a more robust estimate of true parameter values as well as their confidence intervals. For a given parent galaxy sample, $N_j^\mathrm{gal}$ with $N_i^\mathrm{AGN}$ X-ray-detected AGNs in each redshift bin, the extended MXL value of $\pledd$ is given by: 

\begin{equation}
\ln \mathcal{L}= -\mathcal{N} + \sum_{k=1}^{N_i^\mathrm{AGN}} \ln p_k.
\label{equ:logll}
\end{equation}
$p_k$ is the probability that a galaxy of stellar mass $\mathcal{M}_k$ will host an AGN of Eddington ratio $\lambda_{\mathrm{Edd}_k}$, that is, $p_k=p(\lambda_{\mathrm{Edd}_k} \giv \mathcal{M}_k, z_k)$ (Equation~\ref{equ:pledd_withz}) and $\mathcal{N}$ is the expected number of AGNs out of the parent galaxy sample, given the underlying $\pledd$ distribution and our X-ray AGN selection criteria:

\begin{equation}
  \mathcal{N} = \sum^{N^\mathrm{gal}_i}_{j=1}
           \int^{\lambda_{\mathrm{Edd}_{max}}^j}_{\lambda_{\mathrm{Edd}_{min}}^j} p(\ledd \giv \mathcal{M}_j, z_j) ~ \dd\log\ledd
\label{eq:npred}
\end{equation}
The terms $\lambda_{\mathrm{Edd}_{max}}^j$ and $\lambda_{\mathrm{Edd}_{min}}^j$ are determined by the X-ray luminosity range of our AGN selection  and the X-ray depth at the position of galaxy $N^\mathrm{gal}_i$.  
Given the number of free parameters and potential degeneracy between them, we employ the Markov chain Monte Carlo (MCMC) method to estimate best values (and associated uncertainties) that maximize the likelihood in Equation~\ref{equ:logll}. We use the code ``emcee'' as described in~\cite{Foreman-Mackey:2013} to perform the MCMC sampling of the posterior probability distribution. A flat prior probability distribution for all the three parameters is assumed. An example of the posterior probability distribution function as derived with MCMC is shown in Figure~\ref{Fig:MCMC_corner}. Table~\ref{tab:MXLfit} lists the best-fit values for $A$, $\alpha$, and $\gamma$ for the three galaxy populations, as well as the total galaxy population across the whole redshift range $0.5 < z < 2.5$. We have also tried the same fit by dividing the sample into different redshift bins, which yields similar results but with larger uncertainties due to a smaller number of sample galaxies and AGNs.

Our results show that while $\pledd$ for the total galaxy population shows a moderate evolution with redshift of the form $\pledd$~$\propto (1+z)^{1.8}$, different galaxy populations do not follow the same evolution. Among the three galaxy populations, red galaxies exhibit the strongest redshift evolution, with $\pledd$~$\propto (1+z)^{3.7}$ ($\sim 3\sigma$ deviation from that for the total galaxy population), while blue and green galaxies show much weaker redshift evolution, with $\pledd$~$\propto (1+z)^{1.6}$ and $\pledd$~$\propto (1+z)^{1.5}$, respectively. This is consistent with the observed evolution of the AGN fraction with redshift in Section~\ref{subsec:f_agn_z}. We also find marginal evidence (2$\sigma$ level) of a steeper power-law slope for red galaxies compared with those for green and blue galaxies during the MXL fitting across the whole redshift range, in agreement with that obtained through linear fitting of the binned $\ledd$ distribution (Table~\ref{tab:linefit}). Moreover, despite the fact that the MXL fit has incorporated a redshift evolution term, the power-law slopes determined from the MXL fit are consistent with those obtained through a linear fit of binned $\ledd$ distribution. This is an indication that the observed data are well described by our model.
 
Although the redshift evolution of $\pledd$ is determined only for redshifts $0.5 < z < 2.5$ in this work, it is at least qualitatively consistent with that at $z \sim 0$ by \cite{Kauffmann:2009}. Most notably, the normalization of $\pledd$ for red galaxies decreases dramatically at $z\sim 0$ compared with that at high redshift, while $\pledd$ for blue galaxies does not change significantly across the same redshift range. We note that it is not very clear from this comparison what bias is introduced by either X-ray or optically-selected AGNs. Ideally we should compare our results with a large, homogeneous, X-ray-selected AGN sample in the local universe, which is unfortunately not available at present.
Nevertheless, our results at $z \sim 0.5 -2.5$, together with previous work at $z \sim 0$, suggest a rapid evolution of $p (\lambda_{\mathrm{Edd}})$ with redshift for red galaxies and a much slower evolution for blue galaxies. Nevertheless, the shape of $p (\lambda_{\mathrm{Edd}})$ appears relatively constant, suggesting that for each galaxy population, the same physical mechanism is likely to govern the SMBH accretion across cosmic time. In the following section, we explore the underlying physics leading to the aforementioned observational results.

\section{The physical origin of the distinct AGN fraction and Eddington ratio distribution in different hosts}

Here, we first summarize the main results of both the AGN fraction and  $\lambda_{\mathrm{Edd}}$ distribution for different hosts, and then we discuss their possible physical origins. Our results supports the following three major finding.\\
\indent\indent1. The $\lambda_{\mathrm{Edd}}$ distribution for AGNs in the total galaxy population can be well characterized by a power-law with a moderate evolution with redshift, $p(\lambda_{\mathrm{Edd}})~d\log \lambda_{\mathrm{Edd}}  \sim \lambda_{\mathrm{Edd}}^{-0.4}(1+z)^{1.8}~d\log \lambda_{\mathrm{Edd}}$. However, both the power-law slope and redshift evolution differ between sub-populations with different colors. \\
\indent\indent2. The $\lambda_{\mathrm{Edd}}$ distribution for AGNs in red hosts show a power-law distribution at both $z \sim 2$ and $z \sim 1$ with a slope of $\sim -0.6$, which is consistent with what was found for red AGN hosts in the local Universe. While the shape of $\pledd$ remains relatively unchanged, the AGN fraction at fixed $\ledd$ in red galaxies decreases significantly with cosmic time, with $\pledd$~$\propto (1+z)^{3.7}$, leading to a relatively low AGN fraction at low redshift. \\
\indent\indent3. The $\lambda_{\mathrm{Edd}}$ distribution for AGNs in green and blue hosts tends to show a shallower power-law slope and less rapid redshift evolution compared with red hosts. 
The AGN fraction in green galaxies is highest at both redshifts, which is most pronounced at $z \sim 2$ with an enhancement of  high-$\lambda_{\mathrm{Edd}}$ AGNs compared to blue and red galaxies.\\

\subsection{A universal Eddington ratio distribution in massive galaxies across $0.5 < z < 2.5$}
Our results reveal that the Eddington ratio distribution in massive galaxies ($M_{*} > 10^{10} M_{\odot}$) across $0.5 < z < 2.5$ and its redshift evolution can be well characterized by a simple form of $p(\lambda_{\mathrm{Edd}})~d\log \lambda_{\mathrm{Edd}}  \sim \lambda_{\mathrm{Edd}}^{-\alpha}(1+z)^{\gamma}~d\log \lambda_{\mathrm{Edd}}$ with $\alpha \sim 0.4$ and $\gamma \sim 1.8$. This is consistent with several recent studies on low-redshift galaxies~\citep{Aird:2012,Jones:2016} and extends to much higher redshift. We note that some earlier works also unveiled similar results at higher redshifts~\citep{Bongiorno:2012}, but are mainly limited to high-Eddington ratio and high-luminosity AGNs due to the limited depth of X-ray surveys. ~\cite{Bongiorno:2012} found that the redshift evolution of the incidence of AGNs (with typical Eddington ratio of $\ledd \sim 0.1$) follows $\sim (1+z)^{4}$, which may suggest a more rapid evolution for the most extreme AGNs. Though our AGN samples have, in general, lower Eddington ratios, the AGN subsamples with the highest-$\ledd$ also exhibit a more rapid evolution with redshift, as can be seen from the $\pledd$ distribution for the total galaxy population in Figure~\ref{Fig:ledd_dis_z}. 

Our best-fit of the power-law slope of $\pledd$ distribution, $\alpha \sim0.4$, is also in good agreement with these low-redshift studies, within uncertainties. While this simple formula can be directly used to characterize and predict X-ray AGN properties in the total galaxy population, its interpretation  is more complicated. This is because this formula arises from a combined distribution of different $\pledd$ inherent to different sub-populations. Hence galaxies with different colors may not follow the same formula (as we show in this paper, they do not). This means that to gain further insight into the physical origin of the $\ledd$ distribution, it is necessary to study different types of AGN hosts/galaxies separately instead of treating them as a whole. 

\subsection{The high AGN fraction in $z \sim 2$ red galaxies and its rapid decrease with cosmic time}
The significantly low AGN fraction in red galaxies at $z \lesssim 1$ is consistent with the general notion that red galaxies are lacking in cold molecular gas, which is at least necessary (though maybe not sufficient) to fuel an AGN. However, the origin of its significant increase with increasing redshift and the relatively high AGN fraction (compared to blue galaxies) at $z \sim 2$ is unclear. One possible cause could be that the amount of gas reservoir in red galaxies has been relatively high in the past, at $z \sim 2$, when they are on average younger and mostly just formed/quenched. A likely source of the gas reservoir is then the stellar mass loss from an evolving nuclear starburst \citep{Norman:1988}, as also suggested for the power-law $\lambda_{\mathrm{Edd}}$ distribution of AGNs in red galaxies at $z \sim 0$ \citep{Kauffmann:2009}. If the SMBH growth simply traced the stellar mass loss (with no feedback), which decays approximately as $t^{-1}$ to $t^{-1.5}$, then it would lead to a similar power-law distribution of $\lambda_{\mathrm{Edd}}$ with a slope $\alpha$ ranging from $\sim -0.7$ to $\sim -0.9$~\citep{Hogg:1997,Kauffmann:2009}. This is close to, albeit steeper than, what we observe for the power-law slope of the $\lambda_{\mathrm{Edd}}$ distribution at high redshift, and closer to what was found at $z \sim 0$, which is $\sim -0.7$ as shown in~\cite{Kauffmann:2009}. A plausible explanation is that high-redshift red galaxies could maintain some fresh gas in addition to that from the stellar mass loss, especially for those which have only ceased their star formation very recently, as reflected in the low-level star-formation activities observed in $z \sim 2$ passive galaxies \citep{Kriek:2009a,vanDokkum:2010b}. The presence of a significant amount of molecular gas in recently-quenched passive galaxies is also expected in some simulations in which morphologies play a role in quenching their star formation, that is, morphological quenching~\citep{Martig:2009}. These galaxies would tend to host an AGN with relatively high $\lambda_{\mathrm{Edd}}$, which would lead to a shallower power-law slope of the $\lambda_{\mathrm{Edd}}$ distribution than that predicted based on solely mass-loss models. Future gas fraction measurements of $z\sim2$ red AGN hosts, for example, through the Atacama Large Millimeter/submillimeter Array (ALMA), are required to confirm whether or not this is the case.  

However, even if $z \sim 2$ red galaxies indeed have (on average) higher gas fractions than those at $z \sim 1$, it is still not enough to explain the fact that $z \sim 2$ red galaxies have an even higher AGN fraction than blue galaxies, for which the observed gas fraction is much higher ($\sim$ 40-50\%, \citealt{Daddi:2010a,Tacconi:2013}). Therefore, other mechanisms/properties in the $z \sim 2$ red galaxies must facilitate the triggering and  fueling of AGNs. One possibility is that these AGNs are low-luminosity/low-$\lambda_{\mathrm{Edd}}$ relics of more energetic AGNs in their massive star-forming progenitors (or quasars). 
Recent observations detect ubiquitous powerful nuclear outflows in massive star-forming galaxies at $z \sim 2$ \citep{Harrison:2012,Forster:2014}; the descendants of these galaxies, once star formation is quenched, could maintain low-level nuclear activities, as suggested in theoretical models \citep{Hopkins:2009_lifetime}. In this scenario, red galaxies are expected to have gone through a high-$\lambda_{\mathrm{Edd}}$ phase during their latest formation. 
The high duty cycle of high-$\lambda_{\mathrm{Edd}}$ AGNs in green galaxies supports this argument.  

Despite its unknown physical origin, the relatively high AGN fraction in $z \sim 2$ red galaxies provides a promising energy source of "maintenance mode" feedback, which is needed to destroy residual molecular gas reservoirs (if there are any) and suppress further star formation in quiescent galaxies. This is particularly important at high redshift when the typical halo mass of our sample galaxies is not massive enough to fully shut off cold gas accretion~\citep{Dekel:2006}. We note that similar scenarios have been discussed for a sample of optically-selected AGNs in elliptical galaxies at $z\sim 0$~\citep{Schawinski:2009a}. These scenarios are quite different from the conventional maintenance mode feedback from radio AGNs~\citep{Croton:2006}. In contrast to the much rarer radio AGN phenomenon, the high duty cycle of moderate-luminosity X-ray AGNs indicates that most quiescent galaxies at $z \sim 2$ may be able to maintain their quiescence simply through X-ray AGN activities.

In addition to the high probability of hosting an AGN in red galaxies at $z \sim 2$, we also find that the AGNs hosted by red galaxies make up a non-negligible fraction of all the AGNs at both $z \sim 1$ ($\sim$ 20\%) and $z \sim 2$ ($\sim 25\%$). While this high contribution at $z \sim 2$ is driven by the high probability of hosting an AGN in red galaxies, at $z \sim 1$ it is mainly driven by the relative large (compared to blue and green galaxies) number of red galaxies (despite their lower probability of hosting an AGN), as shown in Figure~\ref{Fig:cmd_hist}. Nevertheless, this non-negligible contribution of AGNs in red galaxies must be taken into account when making general comparisons on star-formation properties of AGN-hosts and non-AGN galaxies~\citep[see, e.g., ][]{Mullaney:2015}, especially at high redshifts.

 \begin{figure}[t!]
\begin{center}
\includegraphics[trim= 0 0 0 0,angle=0,width=8cm]{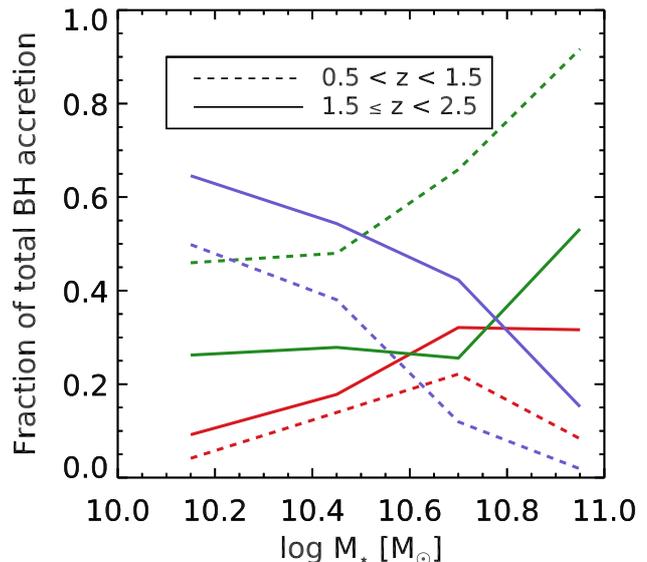}
\caption{The fraction of the total BH mass accretion at a given stellar mass occurring in different galaxy types. Red, green, and blue lines denote BH accretion in the red, green, and blue AGN hosts, respectively. The BH mass accretion in each population of hosts was derived from integrated bolometric (scales as X-ray) luminosity. In general, most BH accretion in high-mass galaxies (BHs) takes place in green and red galaxies while blue galaxies start to dominate BH accretion at lower stellar masses ($M_{*} \lesssim 10^{10.6} M_{\odot}$).    \label{Fig:acc_frac_mass}}
\end{center}
\end{figure}

\subsection{The slow evolution of the AGN fraction in blue galaxies over cosmic time} 
For blue galaxies, we observe a strikingly slow evolution of the $\lambda_{\mathrm{Edd}}$ distribution between $z \sim 1$ and $z \sim 2$, and even $z \sim 0$, despite the fact that the specific star-formation rate or gas fraction increases by a factor of $\sim 5$ from $z \sim 0$ to $z \sim 2$ \citep[see e.g.,][]{Daddi:2007,Tacconi:2013}. This is also consistent with the very weak evolution in the AGN fraction among blue hosts (Figure~\ref{Fig:p_lx_separate}), and with the similarly weak evolution  found for the AGN fraction as a function of SFR between $z\sim1$ and $z\sim0$ \citep{Juneau:2013}.
Therefore, the accretion onto the SMBHs in blue galaxies is clearly not limited by the supply of cold gas. In the local Universe, based on the invariant shape of $\lambda_{\mathrm{Edd}}$ in blue galaxies across a wide range of SMBH mass and stellar ages, \cite{Kauffmann:2009} yield similar arguments. 

Several studies have also shown that above a certain star-formation level, that is, on the main sequence of star-forming galaxies, there is no enhancement of AGN activity with additional star formation \citep{Juneau:2013,Rosario:2013b,Trump:2015}.  This is consistent with our finding that among blue galaxies, the AGN fraction does not change significantly with redshift albeit that the gas fraction or the star-formation rate may change substantially. The flat relation between SFR and $L_{X}$ for star-forming galaxies across $0 < z < 3$, shown in many recent works, also supports this argument \citep{Shao:2010,Rosario:2012,Stanley:2015}.

The slow evolution of both the normalization and the shape of the  $\lambda_{\mathrm{Edd}}$ distribution of blue galaxies also indicates that most of the SMBH growth in blue galaxies may take place in similar host galaxies and does not support a strong redshift evolution of host galaxies, for example, being more merger-driven at higher redshift. Based on high-resolution hydrodynamic simulations, \cite{Gabor:2013} show that isolated gas-rich disks without mergers at $z \sim 2$ typically spend 10-15 per cent of the time detectable as an X-ray AGN (with $L_{X} > 10^{42}$ erg s$^{-1}$), which is in good agreement with our observed value in blue galaxies, $\sim 16\%$.

\subsection{The relatively high AGN fraction in green galaxies} 

For green galaxies, the high AGN fraction (especially at $z=2$) makes it unlikely that their BH accretion is merely driven by a combination of the two accretion modes taking place in red and blue galaxies. Instead, some properties intrinsic to green galaxies must promote a higher duty cycle of AGNs with relatively high $\lambda_{\mathrm{Edd}}$ ($\gtrsim 10^{-2}$), which is particularly true at $z \sim 2$, as shown in the right panel of Figures~\ref{Fig:ledd_dis} and ~\ref{Fig:ledd_dis_z}. As is mentioned for red galaxies, this is also probably required to promote a higher AGN fraction in red galaxies at $z \sim 2$. Most likely, the physical mechanisms leading to a higher AGN duty cycle are also linked to the quenching of star formation and the transformation from blue to red galaxies. A more detailed analysis combining the structural, morphological, and star-formation properties of both AGN-hosts and non-AGN galaxies will be key to exploring this question, which we defer to future work. Here, we speculate that this may be related to recent findings that most progenitors of $z \sim 2$ quiescent (red) galaxies likely belong to a population of compact star-forming galaxies with a high AGN fraction \citep{Barro:2013}.      

Finally, while the $\ledd$ only reflects the current state of SMBH growth in galaxies, by combining galaxies at different masses and redshifts, we can also provide some insight into their growth history. Figure~\ref{Fig:acc_frac_mass} shows the relative contribution to the cosmic SMBH growth from different hosts as a function of stellar mass. The SMBH mass accretion is measured by the integrated X-ray luminosity at each mass bin. Figure~\ref{Fig:acc_frac_mass} reveals that at both redshifts, SMBH mass accretion in blue galaxies tends to dominate at low stellar mass, but decreases as stellar mass increases. On the other hand,  the contribution from red and green hosts to the SMBH mass accretion increases as stellar mass increases, and dominates at the high mass end. This is consistent with the ``downsizing'' evolution of galaxies.

\section{Discussion}
\subsection{Comparison with previous measurements of the Eddington ratio distribution}
In this work, by characterizing galaxies by their extinction-corrected rest-frame $U - V$ colors, we reveal that both the AGN fraction and accretion rates, as well as their evolution with redshift, are closely related to their host colors. These differences would be hidden when considering AGN hosts as a whole or would be obscured when classifying galaxies simply by their observed colors due to, for example, the contamination of a significant population of dusty galaxies ($24~\mu m$ detections) in the red galaxies classified based on observed $U - V$ colors (Figure~\ref{Fig:uv_lmass}), especially at $z\sim 2$. In this section, 
we compare our results with previous studies, and try to reconcile discrepancies between different works at high redshift, and make fair comparisons to studies at low redshift.

In general, our results on the power-law distribution of $\ledd$ for the red galaxies and also the total galaxy population are consistent with previous results at low redshifts~\citep{Kauffmann:2009,Aird:2012}, and extending to higher redshifts. In particular, our best-fit of the power-law slope of $\pledd$ for red galaxies agrees relatively well with those measured by \cite{Kauffmann:2009} and \cite{Aird:2012}. On the other hand, the power-law slope for the total galaxy population, $\alpha \sim 0.4$, is slightly shallower than that measured by \cite{Aird:2012}, which is $\alpha \sim 0.6$. This shallower slope is caused by two factors: the higher fraction of high-$\ledd$ AGNs in blue and green hosts, as also reflected in their shallower slopes of $\pledd$, and the fact that the occupation fraction of blue and green galaxies increases with increasing redshift, hence $\pledd$ for the total galaxy population will more resemble that in blue and green galaxies at higher redshifts.

As for the shape of the $\lambda_{\mathrm{Edd}}$ distribution in blue galaxies, there is some debate in the literature. In the local Universe, \cite{Kauffmann:2009} show that AGNs in star-forming galaxies form a lognormal distribution while those in quiescent galaxies form a power-law distribution. On the other hand, recent work by \cite{Aird:2012} finds a universal power-law shape of $\lambda_{\mathrm{Edd}}$ distribution at $z=0.2-1$, independent of host colors. Our work shows that in the binned $\ledd$ distribution for blue galaxies, there is tentative evidence suggestive of a lognormal distribution of $\pledd$ centered at $\ledd$$\sim 10^{-2}$, like that observed at $z \sim 0$~\citep{Kauffmann:2009}, instead of a power-law. Indeed, the best-fit $\chi^{2}$ value from a lognormal fitting appears slightly smaller than a power-law fit. However, this fit is very sensitive to the lowest $\ledd$ bin, which also has the poorest statistics. A MXL fitting of the unbinned data strongly disfavors such a lognormal distribution (The MXL fitting yields a relatively low mean of the lognormal distribution of $\mu\sim-3$ with a relatively large standard deviation $\sigma \sim 3$, which means that it is essentially not different from a power-law slope in the $\ledd$ range probed by our sample). We hence argue that at least at the $\ledd$ regime spanned by our sample, a simple power-law distribution of $\pledd$ is in good agreement with observations. A power-law distribution is consistent with a number of recent studies of optically-selected AGNs in blue galaxies at lower redshift suggests that the lack of low-$\ledd$ AGNs in blue galaxies is a selection effect caused by the difficulties in selecting weak (low $\ledd$) AGNs from star-forming galaxies with emission line diagnostics~\citep{Trump:2015,Jones:2016}. After correcting for this effect, \cite{Jones:2016} derived an intrinsic power-law slope of $\alpha \sim 0.4$, which is in good agreement with our estimates, though we caution that more direct measurements of $\ledd$ at the low-$\ledd$ regime are essential to further confirmation of this.

\subsection{Possible effects of different morphologies on the Eddington ratio distribution estimates}
As mentioned in Section~\ref{subsec:Ledd}, we chose to use stellar mass as a proxy of SMBH mass instead of bulge mass when estimating $\ledd$. This is mainly driven by the fact that there is no consensus on which one is more tightly correlated with their SMBH masses at high redshifts, and also by the difficulties in accurately measuring bulge masses for individual galaxies at high redshifts. On the other hand, statistical studies do show that blue/green galaxies are less concentrated~\citep[see, e.g.,][]{WangT:2012} and show a smaller bulge-to-total ratio across $0 < z < 2.5$~\citep{Lang:2014}. Therefore, it is interesting to elucidate the possible effects of different morphologies (bulge-to-total ratios) on their $\ledd$ estimation, assuming that bulge mass is a better proxy for SMBH mass.
 
Given the smaller bulge-to-total ratio of blue/green galaxies, they would have a smaller SMBH mass and, hence, a higher $\ledd$ if using bulge mass instead of total stellar mass as an approximation of SMBH mass. This would lead to an even higher fraction of high-$\ledd$ AGNs in green and blue galaxies, which would strengthen our conclusion that green and blue galaxies tend to have a higher abundance of high-$\ledd$ AGNs and shallower power-law slope in $\pledd$. Therefore, we argue that our main conclusion on the differences in the $\ledd$ distribution between different populations remains unchanged no matter what indicators of SMBH masses are used. The median bulge-to-total ratio for quiescent and star-forming galaxies is $\sim 0.25$ and $\sim 0.5$ at $M_{*} \sim 10^{10.6} M_{\odot}$, respectively~\citep{Lang:2014}, with a very weak redshift dependence across $0.5 < z < 2.5$. As a crude estimate, this suggests that the $\ledd$ may be underestimated by a factor of 2 and 4 for red and blue/green galaxies, respectively, if high-redshift galaxies follow the same $M_{\mathrm{BH}}-M_{\mathrm{bulge}}$ relation as local galaxies. Again, we stress that it is at present unclear whether or not this is true at high redshift.
\section{Summary}
In this paper, we explore the dependence of moderate-luminosity AGN activity on extinction-corrected color at 0.5 $< z < 2.5$. By dividing the sample into red, green, and blue galaxies, we reveal that AGN activity varies with host galaxy properties, as is also seen in nearby galaxies. We summarize our main results as follows:

1. X-ray-selected moderate-luminosity AGNs at $z = 0.5-2.5$ are mainly hosted in massive galaxies that span a wide range of host colors. While the majority of them are hosted by blue and green (star-forming) galaxies, a non-negligible fraction of them, $\sim 20\%$-$25\%$ are in truly red galaxies. 

2. For galaxies of all colors, the fraction of galaxies hosting an AGN increases with increasing redshift and stellar mass. However, both the AGN fraction at fixed stellar mass and its evolution with redshift are clearly dependent on host color: while the AGN fraction in red galaxies is significantly lower than that in blue and green galaxies at $z \sim 1$, it increases by a factor of $\sim 5$ at $z \sim 2$, becoming even higher than that in the blue galaxies, which remains nearly unchanged between $z \sim 1$ and $z \sim 2$.  

3. Green galaxies show the highest AGN fraction at all redshifts (which is most pronounced at $z \sim 2$), a factor of 2-3 higher than that in blue galaxies. In particular, among the most massive green galaxies with $M_{*}  \gtrsim 10^{10.6} M_{\odot}$, $\sim 50\%$ of them host an X-ray AGN in our sample. This enhancement is less prominent at $z \sim 1$. The high duty cycle of AGNs in both green and red galaxies at $z \sim 2$ indicates that (X-ray) AGNs could play a role in both quenching star formation and  subsequently maintaining their quiescence at high redshifts.    

4. We show that the Eddington ratio distribution, $p(\lambda_{\mathrm{Edd}})$, for the total galaxy population tends to be a power-law with a moderate redshift evolution in the normalization across $0.5 < z < 2.5$, with $p(\lambda_{\mathrm{Edd}})~d\log \lambda_{\mathrm{Edd}}  \sim \lambda_{\mathrm{Edd}}^{-0.4}(1+z)^{1.8}~d\log \lambda_{\mathrm{Edd}}$. However, $\pledd$, for different populations, differs in term of redshift evolution, with red galaxies showing more rapid redshift evolution than blue or green galaxies. Marginal evidence is also seen for a steeper power-law slope of red galaxies than for blue or green galaxies, though larger samples are required for confirmation of this.

These results suggest that SMBH accretion has a strong dependence on host galaxy color, and studying the AGN and host galaxy population as a whole may lead to ambiguous conclusions. We note that our work still suffers from small number statistics; future X-ray surveys in larger fields with similar sensitivity to GOODS-South are needed to place stronger constraints on the connection between AGN and host galaxies at high redshift.

\begin{acknowledgements}

We thank the anonymous referee for numerous comments and suggestions that clearly improved the consistency and overall quality of this paper.
This work is based on observations taken by the CANDELS Multi-Cycle Treasury Program and the 3D-HST Treasury Program with the NASA/ESA {\it HST}, which is operated by the Association of Universities for Research in Astronomy, Inc., under NASA contract NAS5-26555. T.W. acknowledges support for this work from the National Natural Science Foundation of China under grants 11303014, 11273015 and 11133001. The research leading to these results has received funding from the European Union Seventh Framework Programme (FP7/2007-2013) under grant agreement No. 312725 (ASTRODEEP). YQX acknowledges support of the Thousand Young Talents program (KJ2030220004), the 973 Program (2015CB857004), the CAS Frontier Science Key Research Program (QYZDJ-SSW-SLH006), the National Natural Science Foundation of China (NSFC-11473026, 11421303), the Strategic Priority Research Program ``The Emergence of Cosmological Structures'' of the Chinese Academy of Sciences (XDB09000000), and the Fundamental Research Funds for the Central Universities (WK3440000001).
\end{acknowledgements}
\bibliographystyle{aa}
\bibliography{iero.bib}

\end{document}